\documentclass{conm-p-l}

\copyrightinfo{}{}

\usepackage{graphicx}
\usepackage{subfigure}
 \usepackage{mathrsfs}
\usepackage{amssymb,amsmath,amsthm,amscd}
\usepackage{needspace}

\newtheorem{theorem}{Theorem}[section]
\newtheorem{lemma}[theorem]{Lemma}
\newtheorem{corollary}[theorem]{Corollary}
\newtheorem{proposition}[theorem]{Proposition}
\theoremstyle{definition}
\newtheorem{definition}[theorem]{Definition}
\newtheorem{example}[theorem]{Example}

\newtheorem*{assumption}{\bf Assumptions I}
\theoremstyle{remark}
\newtheorem{remark}[theorem]{Remark}

\numberwithin{equation}{section}

\newcommand{\iii}{{\, \vert\kern-0.25ex\vert\kern-0.25ex\vert\, }}

\DeclareMathOperator{\spn}{span}

\newcommand{\QQ}{\mathcal{Q}}

\newcommand{\Z}{\mathbb{Z}}
\newcommand{\N}{\mathbb{N}}
\newcommand{\R}{\mathbb{R}}

\newcommand{\C}{\mathbb{C}}

\newcommand{\Hi}{\mathscr{H}}
\newcommand{\Bi}{\mathscr{B}}

\newcommand{\Gi}{\mathscr{G}}
\newcommand{\dd}{\partial}

\newcommand{\ra}{\rangle}
\newcommand{\la}{\langle}

\newcommand{\G}{\Gamma}

\begin{document}

\title[Simultaneous global exact controllability in projection]{Simultaneous global exact controllability in 
projection of infinite 1D bilinear Schr\"odinger equations}

\author{A. Duca}
\address{Institut Fourier, Universit\'e Grenoble Alpes, 
Gi\`eres, MO 38610, France}
\email{alessandro.duca@univ-grenoble-alpes.fr}
\urladdr{http://www-fourier.univ-grenoble-alpes.fr/~ducaal/}

\thanks{The author thanks Thomas Chambrion for suggesting him the problem and Nabile Boussa\"id for the periodic 
discussions. He is also grateful to Morgan Morancey for the explanation about the works \cite{morgane1} and 
\cite{morganerse2}.}

\date{}

\dedicatory{}


\begin{abstract}The aim of this work is to study the controllability of infinite bilinear Schr\"odinger equations 
on a segment. We consider the equations \eqref{main_equation} $i\dd_t\psi^{j}=-\Delta\psi^j+u(t)B\psi^j$ in the 
Hilbert 
space $L^2((0,1),\C)$ for every $j\in\N^*$. The Laplacian $-\Delta$ is equipped with Dirichlet homogeneous 
boundary 
conditions, $B$ is a bounded symmetric operator and $u\in L^2((0,T),\R)$ with $T>0$. We prove the 
simultaneous local 
and global exact controllability of infinite \eqref{main_equation} in projection. The local 
controllability is guaranteed for any positive time and we provide explicit examples of $B$ for which our
theory is valid. In addition, we show that the controllability of infinite \eqref{main_equation} in projection onto
suitable finite dimensional spaces is 
equivalent to the controllability of a finite number of \eqref{main_equation} (without projecting).
In conclusion, we rephrase our controllability results in terms of density matrices.
\end{abstract}

\maketitle
\tableofcontents

\section{Introduction}
\subsection{The problem}
In this work, we consider infinite particles constrained in a one-dimensional bounded region and subjected to an 
external control field. A suitable choice for such setting is to model the dynamics of these particles by 
infinitely many bilinear Schr\"odinger equations in the Hilbert space $\Hi=L^2((0,1),\C)$
\begin{equation}\label{main_equation}\tag{BSE}\begin{split}
\begin{cases}
i\dd_t\psi_j(t)=A\psi_j(t)+u(t)B\psi_j(t),\ \ \ \ \ \ \ \ \ & t\in(0,T),\ T>0,\\
\psi_j(0)=\psi_j^0\in L^2((0,1),\C),\ \ &j\in\N^*.\\
\end{cases}
\end{split}
\end{equation}
The Laplacian $A=-\Delta$ is equipped with homogeneous Dirichlet boundary conditions such that 
\[
D(A)=H^2((0,1),\C)\cap H^1_0((0,1),\C).
\]
The bounded symmetric operator $B$ models the action of the external 
field, while the control function $u\in L^2((0,T),\R)$ represents its intensity.

We study the controllability of the infinite bilinear Schr\"odinger equations \eqref{main_equation} at the same 
time $T$, with one unique control $u$ and by projecting onto suitable finite dimensional subspaces of $\Hi$.

In order to detail the purpose of the work, we introduce the following notations. We denote by 
$\G_t^u$ the unitary propagator in $\Hi$ generated by the dynamics of the \eqref{main_equation} in a time 
interval 
$[0,t]$ (when it is defined). Let $\Psi:=(\psi_j)_{j\in\N^*}$ an orthonormal system of $\Hi$. We call 
$\pi_N(\Psi)$ with $N\in\N^*$ the orthogonal projector
\begin{align}\label{projector1}\pi_N(\Psi):\Hi\longrightarrow span{\{\psi_k\ :\ k\leq N\}}.\end{align}
We say that two sequences of functions $(\psi^1_{j})_{j\in\N^*},(\psi^2_{j})_{j\in\N^*}\subseteq 
\Hi$ are unitarily equivalent when there exists $\G\in U(\Hi)$ (the space of the unitary operators in $\Hi$) such 
that 
\[
\psi^1_j=\G\psi^2_j,\ \ \ \ \ \ \ \forall j\in\N^*.
\]

The aim of this work is to study the existence of orthonormal systems $\Psi$ of $\Hi$ so that, for any 
$N\in\N^*$ and for any 
suitable 
$(\psi^1_{j})_{j\in\N^*}$ and $(\psi^2_{j})_{j\in\N^*}$ unitarily equivalent, there exist $T>0$ and $u\in 
L^2((0,T),\R)$ such that
\begin{align}\label{eq1}\pi_N(\Psi)\, \G_T^u\psi^1_j=\pi_N(\Psi)\,\psi^2_j,\ \ \ \ \ \ \ \ \ \forall 
j\in\N^*.\end{align}
If we denote by $\la\cdot,\cdot\ra_{L^2}$ the usual $L^2-$scalar product, then the identities \eqref{eq1} become
\[
\la\psi_k,\G_T^u\psi^1_j\ra_{L^2}=\la\psi_k,\psi^2_j\ra_{L^2},\ \ \ \ \ \ \ \ \ \forall j,k\in\N^*,\ k\leq N.
\]
In order to achieve the result, we show that the simultaneous global exact controllability in projection onto 
suitable $N$ dimensional spaces is equivalent to the controllability of $N$ problems \eqref{main_equation} 
(without 
projecting).

\subsection{Main results}\label{main_results}
Let $\|\cdot\|_{L^2}$ be the norm of the Hilbert space $\Hi=L^2((0,1),\C)$ and $\la\cdot,\cdot\ra_{L^2}$ be the 
corresponding scalar product. Let
\[
(\phi_j)_{j\in\N^*},\ \ \ \ \ \ \  \ \ (\lambda_j)_{j\in\N^*}
\] 
respectively be the eigenfunctions and the 
eigenvalues of $A$ such that
\begin{align}\label{eigenfunctions}\phi_j(x)=\sqrt{2}\sin(j\pi x),\ \ \ \  \ \ \ \ \lambda_j=\pi^2j^2,\ \ \ \  \ \ 
\ \forall j\in\N^*.\end{align}
We notice that $(\phi_j)_{j\in\N^*}$ forms a complete orthonormal system of $\Hi$ and we consider the spaces
\[
H^3_{(0)}:=D(|A|^\frac{3}{2}),\ \ \ \  \ \ \ \  \ 
\|\cdot\|_{(3)}:=\|\cdot\|_{H^3_{(0)}}=\Big(\sum_{k=1}^{\infty}|k^3\la\cdot,\phi_k\ra_{L^2}|^2\Big)^{\frac{1}{2}},
\]
\[
\ell^{\infty}(H^3_{(0)})=\big\{(\psi_j)_{j\in\N^*}\subset{H^3_{(0)}}\big|\ 
\sup_{j\in\N^*}\|\psi_j\|_{(3)}<\infty\big\}.
\]
For $s\in\N^*$, we call $H^s:=H^s((0,1),\C)$, $H_{0}^1:=H_{0}^1((0,1),\C)$ and, for $N\in\N^*$, we define
\begin{equation}\label{I}
I^N:=\{(j,k)\in\N^*\times\{1,\dots,N\}\ :\ j> k\}.
\end{equation}

\begin{assumption}
	The operator $B$ is bounded and symmetric in the Hilbert space $\Hi=L^2((0,1),\C)$. In addition, it 
satisfies the 
following conditions.
	\begin{enumerate}
		\item For any $N\in\N^*$, there exists $C_N>0$ such that $|\la\phi_k,B\phi_j\ra_{L^2}|\geq  
\frac{C_N}{k^{3}}$ for every $j,k\in\N^*$ with $j\leq N$.
		\item $Ran(B|_{H^2_{(0)}})\subseteq H^2_{(0)}$ and $Ran(B|_{H^3_{(0)}})\subseteq H^3\cap H^1_{0}.$
		\item For every $N\in\N^*$ and for every $(j,k),(l,m)\in I^N$ such that $(j,k)\neq (l,m)$ and
$j^2-k^2-l^2+m^2=0$, we have
\[\la\phi_j,B\phi_j\ra_{L^2}-\la\phi_k,B\phi_k\ra_{L^2}-\la\phi_l,B\phi_l\ra_{L^2}+\la\phi_m,B\phi_m\ra_{L^2}\neq 
0.\]
	\end{enumerate}
\end{assumption}
The first condition in Assumptions I quantifies how much $B$ mixes eigenstates, while the second 
fixes 
its regularity. The third condition instead is required in order to decouple, through perturbation theory 
techniques, the eigenvalues resonances appearing in the proof of the following statement.

The next theorem states one of the main results of the work that is the simultaneous global exact controllability 
in projection of infinite \eqref{main_equation}. In order to keep this introduction as simple as possible, we 
postpone to Section \ref{local_control} the second important result of the work which is the simultaneous local 
exact controllability in projection for any positive time (Theorem \ref{theorem_main_local_control}).

\begin{theorem}\label{main_theorem}
	Let $\G_t^u$ be the unitary propagator in $\Hi$ generated by the dynamics of the \eqref{main_equation} in the 
time 
interval $[0,t]$ with $B$ satisfying Assumptions I. Assume that $\Psi:=(\psi_j)_{j\in\N^*}\subset H^3_{(0)}$ is an 
orthonormal system of $\Hi$. Let $(\psi_j^1)_{j\in\N^*}$ and $(\psi_j^2)_{j\in\N^*}\subset H^3_{(0)}$ be 
complete orthonormal systems of $\Hi$ and $\widehat\G\in U(\Hi)$ be the unitary operator such that 
$(\widehat\G\psi_j^2)_{j\in\N^*}=(\psi_j^1)_{j\in\N^*}$. If the following condition is satisfied
\begin{equation}\label{compatibilitycondition1}
(\widehat\G\psi_j)_{j\leq N}\subset H^3_{(0)}\end{equation} 
	with $N\in\N^*$, then there exist $T>0$, $u\in L^2((0,T),\R)$ and $(\theta_k)_{k\leq N}\subset\R$ such that
	\begin{equation}\label{result_projected}\begin{split}
	\la\psi_k,\G^u_T\psi_j^1\ra_{L^2}&=e^{i\theta_k}\la\psi_k,\psi_j^2\ra_{L^2},\ \ \ \ \ \forall j,k\in\N^*,\ 
k\leq N.\\
	\end{split}
	\end{equation}
\end{theorem}

Theorem \ref{main_theorem} allows to control with a single $u$ and at the same time $T$ any finite number of 
components of 
infinitely many solutions of the problems \eqref{main_equation}. We notice that the statement is ensured up to 
phases in the components which prevents to formulate the 
result in terms of projectors. In addition, the 
orthonormal system $(\psi_j)_{j\in\N^*}$ has to verify a $H^3_{(0)}-$compatibility condition exposed in 
\eqref{compatibilitycondition1}. Despite this assumption may seem 
unusual, it spontaneously appears when we try to 
control in projection infinite \eqref{main_equation}. We provide further discussions on the 
subject in Remark \ref{compatibility} where we show that it is a natural constraint for this kind of problems.

When we want to control in projection with respect to the target orthonormal 
system by using Theorem \ref{main_theorem}, we choose $\Psi\equiv\Psi^2$.  In this case, we notice that 
$$(\widehat\G\psi_j)_{j\leq N}=(\widehat\G\psi_j^2)_{j\leq N}=(\psi_j^1)_{j\leq N}\subset H^3_{(0)}$$ and the
$H^3_{(0)}-$compatibility condition 
\eqref{compatibilitycondition1} is trivially satisfied. In addition, 
$$e^{i\theta_k}\la\psi^2_k,\psi_j^2\ra_{L^2}=e^{i\theta_k}\delta_{k,j}=e^{i\theta_j}\la\psi^2_k,\psi_j^2\ra_{L^2},
\ \ \ \ \forall j ,k\in\N^*.$$ Thus, the relations \eqref{result_projected} become
\begin{equation}\begin{split}\begin{cases}\label{projector_equal_target}
\pi_N(\Psi^2)\G^u_T\psi_j^1=\pi_N(\Psi^2)\ e^{i\theta_j}\psi_j^2,\ \ \ \ \ \ \ \ \ \ \ \ \ \  &\forall j\leq N,\\
\pi_N(\Psi^2)\G^u_T\psi_j^1=\pi_N(\Psi^2)\ \psi_j^2,\ \ \ \ \ \ \ \ \ \ \ \ \ \  &\forall j> N.\\
\end{cases}\end{split}\end{equation}
As $\Psi^2$ is composed by orthonormal elements, the projector appearing in the first line of 
\eqref{projector_equal_target} acts 
as the identity operator and the right-hand side of the second line is 
equal to $0$. These facts lead to the following corollary.

\begin{corollary}\label{main_corollary}
	Let $\G_t^u$ be the unitary propagator in $\Hi$ generated by the dynamics of the \eqref{main_equation} in the 
time 
interval $[0,t]$ with $B$ satisfying Assumptions I. Let $\Psi^1:=(\psi_j^1)_{j\in\N^*},$ 
$\Psi^2:=(\psi_j^2)_{j\in\N^*}\subset H^3_{(0)}$ be complete orthonormal systems of $\Hi$. For every $N\in\N^*$, 
there exist $T>0$, $u\in L^2((0,T),\R)$ and $(\theta_j)_{j\leq N}\subset\R$ such that
	\begin{equation*}\begin{split}\begin{cases}
	\G^u_T\psi_j^1=e^{i\theta_j}\psi_j^2,\ \ \ \ \ \ \ \ \ \ \ \ \ \ \ \ & \forall j\leq N,\\
	\pi_{N}(\Psi^2)\ \G^u_T\psi_j^1=0, & \forall j>N.\\
	\end{cases}
	\end{split}
	\end{equation*}
\end{corollary}

Here, one can notice the parallelism between our results with the ones provided in the important work 
\cite{morganerse2} by Morancey and Nersesyan. Indeed, Corollary \ref{main_corollary} implies the controllability 
of 
any finite number of bilinear Schr\"odinger equations when $B$ satisfies Assumptions I. Similar results are 
provided in \cite{morganerse2} and here we rephrase the main one.

\begin{theorem}\cite[Main\ Theorem]{morganerse2}\label{th_morganerse}
Let the bilinear Schr\"odinger equation \eqref{main_equation} be considered with $B=M_\mu$ a multiplication 
operator for a 
function $\mu\in H^4$. Fixed $N\in\N^*$, there exists $\QQ$ a residual set of $H^4$ (an intersection of countably 
many subsets of $H^4$ with dense interiors) such that, for every $B=M_\mu$ with $\mu\in\QQ$, the following result 
is satisfied. For any $(\psi_k^1)_{k\leq N}$, $(\psi_k^2)_{k\leq N}\subset H^3_{(0)}$ unitarily equivalent, 
there exist $T>0$ and $u\in L^2((0,T),\R)$ such that $\G^u_T\psi_k^1=\psi_k^2$ for every $k\leq N.$
\end{theorem}

As we show in Section \ref{provaequivalenza}, the controllability of infinite \eqref{main_equation} in 
projection is 
equivalent to the controllability of a finite number of \eqref{main_equation} (without projecting). In view of 
this 
fact, 
similar statements to Theorem \ref{main_theorem} can be provided by using the theory developed in Section 
\ref{provaequivalenza} and the one from \cite{morganerse2}. Even though such results can be really interesting, 
they are ensured with respect to abstract control operators $B$ (of multiplicative type) and then the 
controllability is 
only generically 
verified. 
From this perspective, our purpose is different. We aim to ensure the 
simultaneous global exact controllability when simple and explicit hypotheses on the problem, such as Assumptions 
I, are satisfied.  This 
fact allows us 
to provide examples of $B$ for which the result is guaranteed, {\it i.e.} $B:\psi\in\Hi\mapsto x^2\psi$ (we 
refer 
to Example \ref{example} for further details on this case and for other examples). Our goal is achieved by using 
different techniques from \cite{morganerse2} whose disadvantage is the loss of control
on the phase terms appearing in Theorem \ref{main_theorem} and Corollary \ref{main_corollary}.

The other main contributions of the work are the following. First, we prove the equivalence between the 
controllability of infinitely many \eqref{main_equation} in projection and the controllability of a finite number 
of equations without projection.  Second, we prove the local controllability in any positive time $T>0$ which is 
stated in Section \ref{local_control}. Third, we use Theorem \ref{main_theorem} and Corollary \ref{main_corollary} 
in order to ensure the global exact 
controllability in projection for density matrices in 
Section 
\ref{density_matrices}.

\subsection{A brief bibliography}
Global approximate controllability results for the bilinear Schr\"odinger equation are provided with different 
techniques in literature. For instance, adiabatic arguments are considered by Boscain, Chittaro, Gauthier, Mason, 
Rossi and Sigalotti in \cite{ugo3} and \cite{ugo2}.
The controllability is achieved with Lyapunov techniques by Mirrahimi in \cite{milo} and by Nersesyan in 
\cite{nerse2}. 
Lie-Galerkin arguments are used by Boscain, Boussa\"id, Caponigro, Chambrion, Mason and Sigalotti in 
\cite{chambrion1}, \cite{nabile} and \cite{chambrion}.

The exact controllability of the bilinear Schr\"odinger equation \eqref{main_equation} is in general a more 
delicate matter as a consequence of the results provided in the work on bilinear systems \cite{ball} by Ball, 
Mardsen and Slemrod. There, they prove that the \eqref{main_equation} is not exactly 
controllable in the Hilbert space where it is defined when $B$ is a bounded operator and $u\in 
L^2_{loc}(\R^+,\R)$ (even though it is well-posed).

Despite this non-controllability result, many authors have addressed the problem for weaker notions of 
controllability by considering suitable subspaces of $D(A)$. This idea was preliminarily introduced by Beauchard 
in 
\cite{beauchard} and popularized by the work in \cite{laurent}. In \cite{laurent}, Beauchard and Laurent 
prove the local exact controllability of \eqref{main_equation} in a neighborhood of the first eigenfunction of $A$ 
in 
$S\cap H^3_{(0)}$ when $B$ is a suitable multiplication operator.
The same kind of operators are considered in \cite{morgane1}, where Morancey 
ensures the simultaneous local exact 
controllability in $S\cap H^3_{(0)}$ for at most three problems \eqref{main_equation} and up to phases. In the 
work \cite{morganerse2}, Morancey and Nersesyan 
extend such 
result and prove Theorem \ref{th_morganerse}. 

\subsection{Scheme of the work}

In Section \ref{Auxiliary_results}, we fix the notations considered in the work and we present some preliminary 
features of 
the problem such as the well-posedness of the \eqref{main_equation} in the space $H^3_{(0)}$ proved in 
\cite{laurent}.

\noindent	
In Section \ref{local_control}, we ensure Theorem \ref{theorem_main_local_control} which states the simultaneous 
local exact controllability in projection for any positive time up to phases. In order to motivate the 
modification of the problem, we emphasize the obstructions to overcome.
	
	\noindent
	In Section \ref{lollol}, we prove Theorem \ref{main_theorem}. First, we 
show that the simultaneous global 
exact 
controllability in projection is equivalent to the controllability of finite \eqref{main_equation} in Proposition 
\ref{equivalenza}. Second, we ensure with Proposition \ref{finite_simultaneous} the simultaneous global exact 
controllability of finite 
\eqref{main_equation} by using the theory from Section \ref{local_control} and a global approximate 
controllability. The propositions 
\ref{equivalenza} and \ref{finite_simultaneous} lead to Theorem 
\ref{main_theorem}.

	\noindent In Section \ref{density_matrices}, we rephrase our results in terms of density matrices, while in 
Section 
\ref{conclu}, we provide some conclusive comments on the work.

	\noindent
    In Appendix \ref{appendix_moment_problem}, we briefly discuss the solvability of the so-called moment 
problems, while in Appendix \ref{appendix_analitics}, we develop the perturbation theory techniques adopted in the 
work.

\section{Auxiliary results}\label{Auxiliary_results}
\subsection{Notations and preliminaries}
We denote by $\Hi$ the Hilbert space $L^2((0,1),\C)$ equipped with the norm $\|\cdot\|_{L^2}$ and the scalar 
product $\la\cdot,\cdot\ra_{L^2}$ such 
that
\[
 \la f,g\ra_{L^2}=\int_0^1\overline{f}(x)g(x)dx,\ \ \ \ \ \ \ \forall f,g\in \Hi. 
 \]
Let $\Bi$ be a Banach space. We introduce for $s>0$, 
\begin{equation}
\begin{split}\label{spaces}
H^s_{(0)}=&D(|A|^\frac{s}{2}),\ \ \ \  \ \ \ \  \ 
\|\cdot\|_{(s)}=\|\cdot\|_{H^s_{(0)}}=\Big(\sum_{k=1}^{\infty}|k^s\la\cdot,\phi_k\ra_{L^2}|^2\Big)^{\frac{1}{2}},\\
&h^s(\Bi)=\Big\{(\psi_j)_{j\in\N^*}\subset{\Bi}\ \big|\ \sum_{j=1}^{\infty}(j^s\|\psi_j\|_{\Bi} 
)^2<\infty\Big\},\\ 
& \ell^\infty(\Bi)=\Big\{(\psi_j)_{j\in\N^*}\subset{\Bi}\ \big|\ 
\sup_{j\in\N^*}\|\psi_j\|_{\Bi} 
<\infty\Big\}.\\
\end{split}
\end{equation}

\noindent
We recall that $(\phi_j)_{j\in\N^*}$ is a complete orthonormal system of $\Hi$ composed by eigenfunctions of $A$ 
defined in \eqref{eigenfunctions} and related to the eigenvalues $(\lambda_j)_{j\in\N^*}$. Fixed
\[
\Psi:=(\psi_j)_{j\in\N^*}\subset\Hi,\ \ \ \ \  \ \ \ \ \Hi_N(\Psi):=\spn \{\psi_j:j\leq N\},
\]
we define 
$\pi_N(\Psi)$ the orthogonal projector such that
\begin{align}\label{projector}\pi_N(\Psi):\Hi\longrightarrow \Hi_N(\Psi).\end{align}

\begin{remark}\label{osss_closed}
	If a bounded operator $B$ satisfies Assumptions I, then $B\in L(H^2_{(0)},H^2_{(0)})$. Indeed, $B$ is closed 
in 
$\Hi$, so for every $(u_n)_{n\in\N^*}\subset\Hi$ such that $u_n\overset{\Hi}{\longrightarrow} u$ and 
$Bu_n\overset{\Hi}{\longrightarrow} v$, we have $Bu=v$.
	Now, for every $(u_n)_{n\in\N^*}\subset H^2_{(0)}$ such that $u_n\overset{H^2_{(0)}}{\longrightarrow} u$ and 
$Bu_n\overset{H^2_{(0)}}{\longrightarrow} v$, the convergences with respect to the $\Hi$-norm are implied and 
$Bu=v$.
	Hence, the operator $B$ is closed in $H^2_{(0)}$ and $B\in L(H^2_{(0)},H^2_{(0)})$. The same argument leads to 
$B\in L(H_{(0)}^3,H^3\cap H^1_{0})$ since $Ran(B|_{H_{(0)}^3})\subseteq H^3\cap H^1_{0}.$
	
\end{remark}
\begin{example}\label{example}
	Assumptions I are satisfied for $B:\psi\mapsto x^2\psi$. Indeed, the condition 1) is guaranteed as
	\begin{equation*}\begin{split}\begin{cases}
	\la\phi_j, x^2\phi_k\ra_{L^2}=
	\frac{(-1)^{j-k}}{(j-k)^2\pi^2}-\frac{(-1)^{j+k}}{(j+k)^2\pi^2},\ \ \ \ \ \ \ & \forall j,k\in\N,\ 
j\neq k,\\
	\la\phi_k, x^2\phi_k\ra_{L^2}=\frac{1}{3}-
	\frac{1}{2k^2\pi^2},\ \ \ &  \forall k\in\N^*.\\
	\end{cases}
	\end{split}
	\end{equation*}
	The point 2) of Assumptions I is trivially true, while the condition 3) is due to the following 
implication. For every $N\in\N^*$ and $(j,k),(l,m)\in I^N$ such that $(j,k)\neq (l,m)$ and 
$j^{2}-k^{2}-l^{2}+m^{2}= 0$, we have 
\[
 j^{-2}-k^{-2}-l^{-2}+m^{-2}\neq 0.
 \]
	We notice that the same properties are valid for other control operators. For instance, if we consider 
$B:\psi\in\Hi\longmapsto \sin\big(\frac{\pi}{2}x\big)\psi$, then Assumptions I are satisfied thanks to the 
identities
	\begin{equation*}\begin{split}\begin{cases}
	\la\phi_j, B\phi_k\ra_{L^2}=-\frac{32 j k}{\pi (16 j^4+16 k^4 -8 j^2- 8k^2- 32 k^2 j^2+1 )}\ \ \ \ \ 
\ \ & \forall j,k\in\N,\ j\neq k,\\
	\la\phi_k, B\phi_k\ra_{L^2}=\frac{2}{\pi}+\frac{2}{(16k^2-1)\pi},\ \ \ &  \forall k\in\N^*.
	\end{cases}
	\end{split}
	\end{equation*} The same is true for the operator $B:\psi\in\Hi\longmapsto x^3\psi.$ Finally, an example of 
operator $B$ satisfying Assumptions I which is not of multiplicative type is
\[
B:\psi\in\Hi\longmapsto \sum_{j\in\N^*}\phi_{2j-1}\big\la\phi_{2j-1}, x^2\psi\big\ra+ 
\sum_{j\in\N^*}\phi_{2j}\Big\la\phi_{2j}, \sin\Big(\frac{\pi}{2}x\Big)\psi\Big\ra\psi.
\] 
\end{example}

\subsection{Well-posedness}\label{well_posed}
In the current subsection, we cite an important result of well-posedness for the following problem in $\Hi$
\begin{equation}\label{main_equation_multiplicationoperator}\begin{split}
\begin{cases}
i\dd_t\psi(t)=A\psi(t)+u(t)\mu\psi(t),\ \ \ \ \ \ \ \ \  \ t\in(0,T),\\
\psi(0)=\psi^0\in L^2((0,1),\C).\\
\end{cases}
\end{split}
\end{equation}
\begin{proposition}\label{proposition_well_posed}\cite[Lemma\ 1;\ 	Proposition\ 2]{laurent}
Let $\mu\in H^3$, $T >0$, $\psi^0\in H^3_{(0)}$ and $u\in L^2((0,T),\R)$. There
exists a unique mild solution of \eqref{main_equation_multiplicationoperator} in
$H^3_{(0)}$, i.e. $\psi\in C^0([0,T],H^3_{(0)})$ so that
\[
\psi(t)=e^{-iA t}\psi^0-i\int_0^t e^{-iA(t-s)}u(s)\mu\psi(s)ds,\ \ \ \ \ \forall t\in[0,T].
\]
Moreover, for every $R>0$, there exists $C=C(T,\mu,R)>0$ such that, if\\ $\|u\|_{L^2((0,T),\R)}<R$, then, for 
every $\psi^0\in H^3_{(0)},$ the solution satisfies 
\[
\|\psi\|_{C^0([0,T],H^3_{(0)})}\leq C\|\psi^0\|_{(3)},\ \ \ \ \  \ 
\ \ \ \|\psi(t)\|_{L^2} =\|\psi^0\|_{L^2},\ \ \ \ \  \ \forall t\in[0,T].
\]
\end{proposition}
\begin{remark}\label{remark_well_posed}The result of Proposition \ref{proposition_well_posed} is not only valid 
for multiplication operators, but also 
for other 
suitable operators $B$. Indeed, the same proofs of \cite[Lemma\ 1]{laurent} and \cite[Proposition\ 2]{laurent} 
lead to the well-posedness of the \eqref{main_equation} when $B$ is a bounded symmetric operator such that 
	\[
B\in L(H^3_{(0)},H^3\cap H^1_0),\ \ \ \ \ \ B\in L(H^2_{(0)}),
\]
	which are verified if $B$ satisfies Assumptions I, thanks to Remark \ref{osss_closed}. \end{remark}
	
	Let 
$\G_t^u$ be the unitary propagator in $\Hi$ generated by the \eqref{main_equation} in the time interval 
$[0,t]$. For any mild solution $\psi_j$ in $L^2((0,1),\C)$ of the $j$-th problem \eqref{main_equation} with 
$j\in\N^*$, we 
have
\[
\G_{t}^{u} \psi_j(0)=\psi_j(t)~.
\]
As a consequence of Remark \ref{remark_well_posed}, it 
follows 
$(\G_T^u\psi_j)_{j\in\N^*}\in \ell^{\infty}(H^3_{(0)})$ for every $(\psi_j)_{j\in\N^*}\in\ell^{\infty}(H^3_{(0)})$. 
 We refer to \eqref{spaces} for the definition of the space
$\ell^{\infty}(H^3_{(0)})$.

\subsection{Time reversibility}\label{time_revers}
An important feature of the bilinear Schr\"odinger equation is the time reversibility. If we consider 
$\psi(t)=\G_t^u\psi^0$ and we substitute $t$ with $T-t$ for $T>0$ in a bilinear Schr\"odinger equation, then we 
have
\begin{equation*}\begin{split}
\begin{cases}
i\dd_t\G_{T-t}^u\psi^0=-A\G_{T-t}^u\psi^0-u(T-t)B\G_{T-t}^u\psi^0,\ \ \ \ \  \ \ \ \ \ t\in (0,T),\\
\G_{T-0}^u\psi^0=\G_{T}^u\psi^0=\psi^1.\\
\end{cases}
\end{split}
\end{equation*}
We define the operator $\widetilde\G_{t}^{\widetilde u}$ such that $\G_{T-t}^{u}\psi^0=\widetilde 
\G_{t}^{\widetilde u}\psi^1$ for $\widetilde u(t):=u(T-t)$ and  
\begin{equation}\label{main_equation_reversed}\begin{split}
\begin{cases}
i\dd_t\widetilde\G_{t}^{\widetilde u}\psi^1=(-A-\widetilde u(t)B)\widetilde\G_{t}^{\widetilde u}\psi^1,\ \ \ \  \ 
\ \ \ \  \ t\in(0,T),\\
\widetilde\G_{0}^{\widetilde u}\psi^1=\psi^1\in L^2((0,1),\C).\\
\end{cases}
\end{split}
\end{equation}
As $\psi^0=\widetilde \G_{T}^{\widetilde u}\G_{T}^{u}\psi^0$ and $\psi^1=\G_{T}^{u}\widetilde \G_{T}^{\widetilde 
u}\psi^1$, it follows $\widetilde\G_T^{\widetilde{u}}=(\G_{T}^u)^{-1}=(\G_{T}^{u})^*.$ The operator 
$\widetilde\G_{t}^{\widetilde u}$ describes the reversed dynamics of $\G_{t}^{u}$ induced by the system 
\eqref{main_equation_reversed} and generated by the Hamiltonian $(-A-\widetilde u(t)B)$.

\section[Simultaneous local controllability]{Simultaneous local exact controllability in 
projection}\label{local_control}
\subsection{Main result}\label{main_result_local_control}
In this section, we examine the simultaneous local exact controllability in projection stated by the following 
theorem.
\begin{theorem}\label{theorem_main_local_control}
	Let $\G_t^u$ be the unitary propagator in $\Hi$ generated by the dynamics of the \eqref{main_equation} in the 
time interval $[0,t]$ with $B$ satisfying Assumptions I. Let $N\in\N^*.$
For every $T>0$, there exist an open set $O$ in $\ell^\infty(H^3_{(0)})$ and an orthonormal 
system $\Psi:=(\psi_j)_{j\in\N^*}\in O$ such that the following result is verified.  Let
$(\psi_j^1)_{j\in\N^*}\in O$ be a complete 
orthonormal 
system and $\widehat\G\in U(\Hi)$ be such that 
$(\widehat\G\psi_j^1)_{j\in\N^*}=(\psi_j)_{j\in\N^*}$. If $(\widehat\G\psi_j)_{j\leq N}\subset 
H^3_{(0)},$ then there exist $(\theta_j)_{j\leq N}\subset\R$ and $u\in L^2((0,T),\R)$ such that
	\begin{equation*}\begin{split}\begin{cases}
\la\psi_k,\G_T^u\psi_j\ra_{L^2}=e^{i\theta_j}\la\psi_k,\psi^1_j\ra_{L^2},\ \ \ \ \ \ \ \ \ &\forall j,k\in\N^*,\ 
j\leq N,\ k\leq N,\\
\la\psi_k,\G_T^u\psi_j\ra_{L^2}=\la\psi_k,\psi^1_j\ra_{L^2},\ \ \ \ \ \ \ \ \ &\forall j,k\in\N^*,\ j>N,\ k\leq 
N.\end{cases}
	\end{split}
	\end{equation*}
	In other words, the following identities are satisfied (with $\pi_N(\Psi)$ defined in \eqref{projector1}):
	\begin{equation*}\begin{split}\begin{cases}
\pi_N(\Psi)\G_T^u\psi_j=e^{i\theta_j}\pi_N(\Psi)\psi^1_j,\ \ \ \ \ \ \ \ \ &\forall j\in\N^*,\ j\leq N,\\
\pi_N(\Psi)\G_T^u\psi_j=\pi_N(\Psi)\psi^1_j,\ \ \ \ \ \ \ \ \ &\forall j\in\N^*,\ j>N.\end{cases}
	\end{split}
	\end{equation*}
	\end{theorem}
	\subsection{Introductive discussion}\label{introductive_discussion}
We start by explaining why we need to modify the problem in order to prove Theorem 
\ref{theorem_main_local_control}. Let 
$\Phi=(\phi_j)_{j\in\N^*}$ be a complete orthonormal system composed by eigenfunctions of $A$. For every 
$j\in\N^*,$ we denote 
$\phi_j(t,x)=e^{-i\lambda_j t}\phi_j(x)$ with $t>0$. From now on, we adopt the notation 
$\phi_j(t)=\phi_j(t,\cdot)$. Let $\epsilon>0$ and $T>0$. We consider the set
\begin{equation}\label{neigh_base}\begin{split}
O_{\epsilon,T}:=\Big\{&(\psi_j)_{j\in \N^*}\in\ell^\infty(H^3_{(0)})\text{ complete orthonormal system of $\Hi$ 
such}\\
&\text{that } \sup_{k\leq 
N}\sum_{j\in\N^*}k^6|\la\psi_j,\phi_k(T)\ra_{L^2}-\la\phi_j(T),\phi_k(T)\ra_{L^2}|^2<\epsilon\Big\}.\\
\end{split}
\end{equation} 
We would like to prove to validity of Theorem \ref{theorem_main_local_control} in the neighborhood 
$O_{\epsilon,T}$ for $\epsilon,T>0$ with respect to 
the projector $\pi_N(\Phi)$ (see the definition \eqref{projector}). Then,
\[
\G_T^u\phi_j=\sum_{k=1}^{\infty}{\phi_k(T)}\la \phi_k(T), \G_{t}^u\phi_j\ra_{L^2},\ \ \  \ \ \ \ \  
\phi_j(T)=e^{-i\lambda_j T}\phi_j,\ \ \ \  \ \ \ \forall j\in\N^*
\]
is the solution of the {j-th} \eqref{main_equation} with initial data $\phi_j$ at time $T>0$. We consider the 
infinite 
matrix $\alpha(u)$ such that \[
\alpha_{k,j}(u)=\la \phi_k(T), \G_{T}^u\phi_j\ra_{L^2},\ \ \ \ \  \ \ \ \forall 
k,j\in\N^*,\ k\leq N.
\]
We would like to ensure the existence of $\epsilon>0$ and $T>0$ such that for any $(\psi_j)_{j\in\N^*}\in 
O_{\epsilon,T}$, there exists $u\in L^2((0,T),\R)$ such that
\[
\pi_N(\Phi)\G_T^{u}\phi_j=\pi_N(\Phi)\psi_j,\ \ \ \ \ \ \forall 
j\in\N^*.
\]
This result can be proved by studying the local surjectivity of $\alpha$ for $T>0$. To this purpose, we want to 
use 
the inverse mapping theorem and study the surjectivity of the Fr\'echet derivative of $\alpha$ the infinite matrix 
$\gamma(v):=(d_u\alpha(0))\cdot\ v$ such that
\begin{equation*}\begin{split}
\gamma_{k,j}(v):&=\left\la\phi_k(T), 
-i\int_0^Te^{-iA(T-s)}v(s)Be^{-iAs}\phi_jds\right\ra_{L^2}\\
&=-i\int_{0}^Tv(s)e^{-i(\lambda_j-\lambda_k)s}ds 
B_{k,j},\ \ \ \ \ \ \ \ \forall j,k\in\N^*,\ k\leq N,\\
\end{split}\end{equation*}
with $B_{k,j}=\la\phi_k,B\phi_j\ra_{L^2}=\la B\phi_k,\phi_j\ra_{L^2}=\overline{B_{j,k}}$. The surjectivity of 
$\gamma$ consists in proving the solvability of the moment problem
\begin{equation}\label{first_moment_problem}\begin{split}
\frac{x_{k,j}}{B_{k,j}}=-i\int_{0}^Tu(s)e^{-i(\lambda_j-\lambda_k)s}ds, \ \ \ \ \ \ \  \ \ \forall j,k\in\N^*,\ 
k\leq N,\\
\end{split}\end{equation}
for each infinite matrix ${\bf x}:=(x_{k,j})_{\underset{k\leq N}{j,k\in\N^*}}$ belonging to a suitable space. To 
this end, one would use Corollary \ref{solvability_moment_problem} which is consequence of the Haraux's Theorem 
but an obstruction 
appears. The terms $(\lambda_j-\lambda_k)_{\underset{k\leq N}{j,k\in\N^*}}$ in the moment problem 
\eqref{first_moment_problem} 
present the so-called eigenvalues resonances. Formally, for some $j,k,n,m\in\N^*$ such that $j\neq 
k$, 
$n\neq m$, $(j,k)\neq(n,m)$ and $k,m\leq N$, there holds $\lambda_j-\lambda_k=\lambda_n-\lambda_m$, which implies
\begin{equation}\label{constr_moment_problem}\begin{split}
\frac{x_{k,j}}{B_{k,j}}&=-i\int_{0}^Tu(s)e^{-i(\lambda_j-\lambda_k)s}ds=-i\int_{0}^Tu(s)e^{-i(\lambda_n-\lambda_m)s
}ds=\frac{x_{n,m}}{B_{n,m}}.\\
\end{split}\end{equation}
An example of eigenvalues resonance is $\lambda_7-\lambda_1=\lambda_8-\lambda_4$, but many others can be listed. 
For instance, all the diagonal terms of $\gamma$ since $\lambda_j-\lambda_k=0$ for $j=k$. The relation 
\eqref{constr_moment_problem} represents a constraint on the considered matrices ${\bf x}$ which is not naturally 
satisfied in our framework.

In order to avoid this phenomenon, we adopt the following strategy. First, we consider the Hamiltonian 
characterizing the bilinear Schr\"odinger equations \eqref{main_equation} and we use the following decomposition 
\begin{align}\label{decompositionHamiltonia}A+u(t)B=(A+u_0B)+u_1(t)B,\ \ \ \ \ \ \ u_0\in\R,\ u_1\in 
L^2((0,T),\R).\end{align}
Second, we consider $A+u_0B$ instead of $A$. We repeat the previous steps by considering 
$(\phi_j^{u_0})_{j\in\N^*}$ a complete orthonormal system of $\Hi$ composed by eigenfunctions of $A+u_0B$ and 
$(\lambda_j^{u_0})_{j\in\N^*}$ the corresponding eigenvalues. By using $u_0B$ as a perturbation in 
$A+u_0B$, 
we modify the eigenvalues gaps
\[
\lambda_j^{u_0}-\lambda_k^{u_0},\ \  \ \ \ \ \  \ \ \forall j,k\in\N^*,\ k\leq N
\]
in order to remove all the non-diagonal resonances.
Afterwards, we consider $\widehat\alpha$ depending on the parameter $u_0$ (instead of $\alpha$) such that it is 
defined by the elements $\widehat\alpha_{k,j}(u)=\la e^{-i\lambda_k^{u_0}T}\phi^{u_0}_k, 
\G_{T}^u\phi_j^{u_0}\ra_{L^2}$ with $ k,j\in\N^*$ and $k\leq N.$
Now, we rotate the terms of $\widehat\alpha$ in order to remove the resonances on the diagonal terms. We denote by 
$\alpha^{u_0}$ the obtained map, which is defined by the elements
\[
\alpha^{u_0}_{k,j}(u)=\frac{\overline{\widehat\alpha_{j,j}(u)}}{|\widehat\alpha_{j,j}(u)|}\widehat\alpha_{k,j}(u)
, \  \ \ \ \ \  \ \ \forall j,k\in\N^*,\ k\leq N.
\]
In conclusion, we use the inverse mapping theorem with respect to the map $\alpha^{u_0}$.

The first step of our strategy is not so different from the techniques leading to \cite[Main\ 
Theorem]{morganerse2}, however it presents an important difference. In our work, we seek for 
explicit conditions on the operator $B$ such that the perturbative argument is 
valid. 
On 
the contrary, the authors of \cite{morganerse2} prove 
the existence of $\QQ$, a residual subset of $H^4((0,1),\R)$ (in the spirit of Theorem 
\ref{th_morganerse}), such that the controllability holds when $B$ is  a 
multiplication operator by a function $\mu\in\QQ$.

\subsection{The modified problem}\label{modified_problem}
In this subsection, we rewrite the \eqref{main_equation} by applying the decomposition 
\eqref{decompositionHamiltonia} and we 
introduce the groundwork required to apply the strategy discussed in Section \ref{introductive_discussion}. Let 
$u(t)=u_0+u_1(t)$ with $u_0\in\R$,  $u_1\in L^2((0,T),\R)$ and $T>0$. We consider the following Cauchy problems
\begin{equation}\label{main_equation_modified}\begin{split}
\begin{cases}
i\dd_t\psi_j(t)=(A+u_0B)\psi_j(t)+u_1(t)B\psi_j(t),\ \ \ \ \ & t\in(0,T)  ,\\
\psi_j^0=\psi_j(0),\ &j\in \N^*.\\
\end{cases}
\end{split}
\end{equation}
As $B$ is bounded, $A+u_0B$ has pure discrete spectrum. We recall that $(\lambda_j^{u_0})_{j\in\N^*}$ are the 
eigenvalues of $A+u_0B$ and $\Phi^{u_0}:=(\phi_j^{u_0})_{j\in\N^*}$ is a complete 
orthonormal system of $\Hi$ made by corresponding eigenfunctions. Fixed $N\in\N^*$, for every $T>0$ and 
$\epsilon_0>0$, we denote 
\begin{equation}\begin{split}\label{neigh_perturbated}
O_{\epsilon_0,T}^{u_0}:=\Big\{&(\psi_j)_{j\in \N^*}\in \ell^\infty(H^3_{(0)})\text{ complete orthonormal system of 
$\Hi$ such}\\
&\text{that}\sup_{k\leq 
N}\sum_{j\in\N^*}k^6|\la\psi_j,\phi_k^{u_0}(T)\ra_{L^2}-\la\phi_j^{u_0}(T),\phi_k^{u_0}(T)\ra_{L^2}
|^2<\epsilon_0\Big\},\\
\end{split}
\end{equation}
with $\phi_j^{u_0}(T):=e^{-i\lambda_j^{u_0}T}\phi_j^{u_0}$. We choose $|u_0|$ small such that $\lambda_k^{u_0}\neq 
0$ 
for every $k\in\N^*$ (Lemma \ref{perturb_eigenvalues}). The modification of the problem imposes to 
define the space 
\[
\widetilde H^3_{(0)}:=D(|A+u_0B|^{\frac{3}{2}}),\ \ \ \  \ \ \ \  \|\cdot\|_{\widetilde 
H^3_{(0)}}=\Big(\sum_{k=1}^{\infty}\big||\lambda_{k}^{u_0}|^\frac{3}{2}\la\cdot,\phi_k^{u_0}\ra_{L^2}\big|^2\Big)^{
\frac{1}{2}}.
\]
However, we consider from now on $u_0$ in the neighborhood provided by Lemma \ref{perturnated_norm_equivalence} 
so that $\widetilde 
H^3_{(0)}\equiv H^3_{(0)}.$
As introduced in Section \ref{introductive_discussion}, we consider the map $\widehat{\alpha}$ with elements 
$\widehat\alpha_{k,j}(u_1)=\la\phi^{u_0}_k(T), \G_{T}^{u_0+u_1}\phi_j^{u_0}\ra_{L^2}$ for $k\leq N$ and 
$j\in\N^*$. 
The map $\alpha^{u_0}$ is the infinite matrix with elements
\begin{equation}\label{defi_alfa}
\begin{split}
&\begin{cases}\alpha_{k,j}^{u_0}(u_1)=\frac{\overline{\widehat\alpha_{j,j}(u_1)}}{|\widehat\alpha_{j,j}(u_1)|}
\widehat\alpha_{k,j}(u_1), \ \  \ \ \ \ \ \ \ \ \ \ \ \ \ \ \ \ &\forall j,k\in\N^*,\ j,k\leq N,\\
\alpha^{u_0}_{k,j}(u_1)=\widehat\alpha_{k,j}(u_1),\ & \forall j,k\in\N^*,\ j>N,\ k\leq N.\\
\end{cases}
\end{split}
\end{equation}
Now, we study the space where $\alpha^{u_0}$ takes value. Let $\widetilde \G_t^{\widetilde u}$ be the 
propagator of the reversed dynamics defined in Section \ref{time_revers} for $t\in[0,T]$, $u\in 
L^2((0,T),\R)$ and $T>0$. For every $k\in\N^*$, $u_0\in\R$ and $u_1\in L^2((0,T),\R)$, from Proposition 
\ref{proposition_well_posed}, Remark \ref{remark_well_posed} and 
Lemma \ref{perturnated_norm_equivalence}, there exists $C>0$ so that
\begin{equation*}\begin{split}
\sum_{j=1}^{+\infty}j^6|\alpha_{k,j}^{u_0}(u_1)|^2&=\sum_{j=1}^{+\infty}j^6|\la\widetilde\G_T^{u_0+\widetilde 
u_1}\phi_k^{u_0},\phi_j^{u_0}\ra_{L^2}|^2=\|\widetilde\G_T^{ {u_0+\widetilde u_1}}\phi^{u_0}_k\|_{\widetilde 
H^3_{(0)}}^2\\
&\leq C\|\widetilde\G_T^{ {u_0+\widetilde u_1}}\phi^{u_0}_k\|_{(3)}^2<\infty.
\end{split}\end{equation*}
Thus, each $(\alpha_{k,j}^{u_0}(u_1))_{j\in\N^*}\in h^3(\C)$ (defined in \eqref{spaces}). For 
every $(\psi_j)_{j\in\N^*}$ such that $(\psi_j)_{j\in\N^*}\in O^{u_0}_{\epsilon_0,T}$ or such that
$(\psi_j)_{j\in\N^*}=(\G_T^{u_0+u_1}\phi_j^{u_0})_{j\in\N^*}$, we have
\begin{equation*}\begin{split}
\delta_{j,k}&=\la\phi^{u_0}_j,\phi^{u_0}_k\ra_{L^2}=\Big\la\sum_{m\in\N^*}\psi_m\la\psi_m,\phi_j^{u_0}\ra_{L^2},
\sum_{l\in\N^*}\psi_l\la\psi_l,\phi_k^{u_0}\ra_{L^2}\Big\ra_{L^2}\\
&=\Big\la(\la\psi_m,\phi_j^{u_0}\ra_{L^2})_{m\in\N^*},(\la\psi_m,\phi_k^{u_0}\ra_{L^2})_{m\in\N^*}\Big\ra_{
\ell^2},\ \ \ \ \ \forall j,k\leq N.
\end{split}\end{equation*}
	The last relations imply that $\alpha^{u_0}:u\in 
L^2((0,T),\R)\longmapsto(\alpha_{k,j}^{u_0}(u))_{\underset{k\leq N}{k,j\in\N^*}}\in Q^N $ where
	\begin{equation*}
	\begin{split}
	Q^N:=\Big\{&(x_{k,j})_{\underset{k\leq N}{k,j\in\N^*}}\in (h^3(\C))^N\big|\ \ \ x_{k,k}\in\R,\\
	&\big\la(x_{j,l})_{{l\in\N^*}},(x_{k,l})_{{l\in\N^*}}\big\ra_{\ell^2}=\delta_{j,k},\ \ \ \ \forall k,j\leq N
	\Big\}.\\
	\end{split}
	\end{equation*}
Now, $\G_T^{(\cdot)}\psi:u\in 
L^2((0,T),\R)\longmapsto \G_T^{u}\psi\in H^3_{(0)}$ with $\psi\in H^3_{(0)}$ is $C^1$ (see 
\cite[Proposition\ 47]{beauchard} or 
\cite[Section 2]{morgane1} for further details) and the same is true for $\widetilde \G_T^{(\cdot)}\psi:u\in 
L^2((0,T),\R)\longmapsto \widetilde \G_T^{u}\psi\in H^3_{(0)}$ for every $\psi\in H^3_{(0)}$. Finally, the map 
$$\alpha^{u_0}:u\in 
L^2((0,T),\R)\longmapsto(\alpha_{k,j}^{u_0}(u))_{\underset{k\leq N}{k,j\in\N^*}}\in Q^N $$ is $C^1$ thanks to the 
identities 
$\widehat\alpha_{k,j}(u_1)=\la\phi_k^{u_0}(T),\G_T^{u_0+u_1}\phi_j^{u_0}\ra=\la\widetilde\G_T^{u_0+,\widetilde u_1} 
\phi_k^{u_0}(T),\phi_j^{u_0}\ra$ for every $k,j\in\N^*$ with $k\leq N.$ We denote by 
$\gamma^{u_0}(v)=((d_{u_1}\alpha^{u_0})(0))\cdot v$ the Fr\'echet derivative of $\alpha^{u_0}$. Defined 
$\widehat\gamma_{k,j}(v)=((d_{u_1}\widehat\alpha)(0))\cdot v$, the elements of $\gamma^{u_0}(v)$ are
\begin{equation*}\begin{split}
\begin{cases}\gamma_{k,j}^{u_0}=\big(\overline{\widehat\gamma_{j,j}}\delta_{k,j}+\widehat\gamma_{k,j}-\delta_{k,j}
\Re(\widehat\gamma_{j,j})\big),\ \ \ \ \ &\forall j,k\in\N^*,\ j,k\leq N,\\
\gamma_{k,j}^{u_0}=\widehat\gamma_{k,j},\ \ &\forall j,k\in\N^*,\  k\leq N,\ 
j>N\end{cases}\end{split}\end{equation*}
and then, for $B_{k,j}^{u_0}= 
\la\phi_k^{u_0},B\phi_j^{u_0}\ra_{L^2}$ for $k\leq N$ and $j\in\N^*$,
\begin{equation}\begin{split}
 \begin{cases}\label{gamma}
\gamma_{k,j}^{u_0}=\widehat\gamma_{k,j}=-i\int_{0}^Tu_1(s)e^{-i(\lambda_j^{u_0}-\lambda_k^{u_0})s}dsB^{u_0}_{k,j},
\ \ \ \ \ & \forall j,k\in\N^*,\ k\neq j,\\
\gamma_{k,k}^{u_0}=\Re(\widehat\gamma_{k,k})=0,& \forall k\in\N^*.\\
\end{cases}\end{split}\end{equation}
The relation $\gamma_{k,k}^{u_0}=0$ is due to the fact that $(i\widehat\gamma_{k,k})\in\R$ since
$\widehat\gamma_{k,j}=-\overline{\widehat\gamma_{j,k}}$ 
for $j,k\leq N.$ 
Hence, the diagonal elements of $\gamma^{u_0}$ are all equal to $0$ due to the rotations adopted in the definition 
$\alpha^{u_0}$. 	Since $O_{\epsilon_0,T}^{u_0}$ is composed by orthonormal elements, the 
tangent space of $O_{\epsilon_0,T}^{u_0}$ in the point ${\Phi^{u_0}}$ is 
	\begin{equation*}\begin{split}
	T_{\Phi^{u_0}}O_{\epsilon_0,T}^{u_0}=&\Big\{(\psi_j)_{j\in \N^*}\subset \ell^\infty(H^3_{(0)})\big|\ \la 
\phi_k^{u_0},\psi_j\ra_{L^2}=-\overline{\la 
		\phi_j^{u_0},\psi_k\ra_{L^2}}\Big\}.\\
	\end{split}\end{equation*}The last relation implies that $\gamma^{u_0}:u\in 
L^2((0,T),\R)\longmapsto(\gamma_{k,j}^{u_0}(u))_{\underset{k\leq N}{k,j\in\N^*}}\in G^N $ where
	\begin{equation*}
	\begin{split}
		G^N:=\big\{&(x_{k,j})_{\underset{k\leq N}{k,j\in\N^*}}\in (h^3(\C))^N\big|\ \ \  
x_{k,j}=-\overline{x_{j,k}},\  \ \ \  x_{k,k}=0,\ \ \ \ \forall k,j\leq N\big\}.\\
	\end{split}
	\end{equation*}

\begin{remark}\label{control_projection_equivalent}
	When the third point of Remark \ref{remark_general_perturbation} is valid, the controllability in 
$O^{u_0}_{\epsilon_0,T}$ 
(defined in \eqref{neigh_perturbated}) with $\epsilon_0>0$ ensures the controllability in $O_{\epsilon,T}$ 
(defined in 
\eqref{neigh_base}) for suitable $\epsilon>0$. 
Let $(\psi_j)_{j\in\N^*}\in O_{\epsilon,T}$ and $\widehat\G\in U(\Hi)$ be such that 
$(\widehat\G\psi_j)_{j\in\N^*}=(\phi^{u_0}_j)_{j\in\N^*}$ and satisfying $(\widehat\G\phi^{u_0}_j)_{j\leq 
N}\subset H^3_{(0)}$. There exists $C>0$ so that, for every $k\leq N$,
\begin{align}\label{local_compatibility}\sum_{j\in\N^*}|j^3\la\phi_k^{u_0},\psi_j\ra_{L^2}|^2=\sum_{j\in\N^*}
|j^3\la\widehat\G\phi_k^ { u_0 } , \phi^ { u_0 } _j\ra_{L^2}|^2\leq 
C\|\widehat\G\phi_k^{u_0}\|_{(3)}<\infty,\end{align}
thanks to Lemma \ref{perturb_eigenvalues} and Lemma \ref{perturnated_norm_equivalence}. Now, fixed 
$e^{i\theta_j^1}:=\frac{\overline{\la\phi_j^{u_0}(T),\psi_j\ra_{L^2}}}{|\la\phi_j^{u_0}(T),\psi_j\ra_{L^2}|}$ for 
$j\leq N$ and $e^{i\theta_j^1}:=1$ for $j>N$, the relation \eqref{local_compatibility} yields that 
$(e^{i\theta_j^1}\la\phi_k^{u_0}(T),\psi_j\ra_{L^2})_{\underset{k\leq N}{j,k\in\N^*}}$ belongs to $Q^N_{\epsilon}$ 
where
\begin{equation*}\begin{split}
Q^N_{\epsilon}:=\{(x_{k,j})_{\underset{k\leq N}{k,j\in\N^*}}\in Q^N\ |\ \sup_{k\leq 
N}\sum_{j\in\N^*}k^6|x_{k,j}-\delta_{k,j}|^2< \epsilon\}.\end{split}\end{equation*} 
When $\alpha^{u_0}$ is surjective in $Q_\epsilon^N$, there exist $T>0$ and $u\in L^2((0,T),\R)$ such that
\begin{equation}\label{rest}\begin{split}(e^{i\theta_j^1}\la\phi_k^{u_0},\psi_j\ra_{L^2})_{\underset{k\leq 
N}{j,k\in\N^*}}=(e^{i\theta_j^2}\la\phi_k^{u_0},\G_T^u\phi_j^{u_0}\ra_{L^2})_{\underset{k\leq 
N}{j,k\in\N^*}},\end{split}\end{equation}
\begin{equation*}\begin{split}\text{with}\ \ \ \ 
\begin{cases}e^{i\theta^2_j}:=\frac{\overline{\widehat\alpha_{j,j}(u_1)}}{|\widehat\alpha_{j,j}(u_1)|},\ \ \ \  \ 
&j\leq N,\\
e^{i\theta^2_j}:=1,\ \ \ \  \ \ \ &j>N.\end{cases}\end{split}\end{equation*}
Thus, the surjectivity of the map $\alpha^{u_0}$ in $Q^N_{\epsilon_0}$ ensures the validity of Theorem 
\ref{theorem_main_local_control} with respect to the projector $\pi_N(\Phi^{u_0})$ in $O_{\epsilon_0,T}^{u_0}$ and 
in 
$O_{\epsilon,T}$ for a suitable $\epsilon>0$.
\end{remark}

\subsection{Proof of Theorem \ref{theorem_main_local_control}}\label{proof_local_control}

In the next proposition, we state the simultaneous local exact controllability in projection for any $T>0$ up to 
phases. The result implies Theorem \ref{theorem_main_local_control}.

\begin{proposition}\label{proposition_local_control}
	Let $N\in\N^*$ and $B$ satisfy Assumptions I. For every $T>0$, there exist $\epsilon>0$ and $u_0\in\R$ such 
that the following result is verified. Let $(\psi_j^1)_{j\in\N^*}\in O_{\epsilon,T}$ (defined in 
\eqref{neigh_base}) 
and $\widehat\G\in U(\Hi)$ be such that $(\widehat\G\psi_j^1)_{j\in\N^*}=(\phi^{u_0}_j)_{j\in\N^*}$. If 
$(\widehat\G\phi^{u_0}_j)_{j\leq N}\subset H^3_{(0)},$ then there exist
$(\theta_j)_{j\leq N}$ and $u\in L^2((0,T),\R)$ such 
that	\begin{equation*}\begin{split}\begin{cases}
\pi_N(\Phi^{u_0})\G_T^u\phi^{u_0}_j=e^{i\theta_j}\pi_N(\Phi^{u_0})\psi^1_j,\ \ \ \ \ \ \ \ \ &\forall j\in\N^*,\ 
j\leq 
N,\\
\pi_N(\Phi^{u_0})\G_T^u\phi^{u_0}_j=\pi_N(\Phi^{u_0})\psi^1_j,\ \ \ \ \ \ \ \ \ &\forall j\in\N^*,\ j>N.\end{cases}
	\end{split}
	\end{equation*}
\end{proposition} 
\begin{proof} 
	{\bf 1)} Let $u_0$ belong to the neighborhoods defined in Appendix \ref{appendix_analitics} by Lemma 
\ref{perturb_eigenvalues}, Lemma \ref{perturbation_mixing_eigenspaces}, Lemma \ref{perturnated_norm_equivalence} 
and Remark \ref{remark_general_perturbation}. As discussed in Remark \ref{control_projection_equivalent}, the 
surjectivity in $Q^N_\epsilon$ of the map $\alpha^{u_0}$ guarantees the simultaneous local exact controllability 
in projection up to phases in $O_{\epsilon,T}$.

	We want to use the inverse mapping theorem by considering that $G^N$ is the tangent space of $Q^N$ in the 
point 
$(\delta_{k,j})_{\underset{k\leq N}{k,j\in\N^*}}=\alpha^{u_0}(0)$. If $\gamma^{u_0}$ is surjective in $G^N$ for 
$T>0$, then $\alpha^{u_0}$ is surjective in $Q^N_\epsilon$ for $\epsilon$ small enough. The surjectivity of 
$\gamma^{u_0}$ corresponds to the solvability of the moment problem
	\begin{equation}\begin{split}\label{moment_problem_second}
	{x_{k,j}^{u_0}}/{B^{u_0}_{k,j}}=-i\int_{0}^Tu(s)e^{-i(\lambda_j^{u_0}-\lambda_k^{u_0})s} ds,\ \ \ \ \ \ 
\forall 
j,k\in\N^*,\ k\leq N\\
	\end{split}\end{equation}
	for every $\big(x_{k,j}^{u_0}\big)_{\underset{k\leq N}{j,k\in\N^*}}\in G^N$. We notice that the equations 
\eqref{moment_problem_second} for $k=j$ are redundant as $\gamma_{k,k}^{u_0}=0$ and $x_{k,k}^{u_0}=0$ for every 
$k\leq N$ 
since $(x^{u_0}_{k,j})_{\overset{k,j\in\N^*}{k\leq N}}\in G^N$
	. The same is true for $j,k\leq N$ such that $j< k$ since 
\[
(x_{j,k})_{j,k\leq N},\ \ \ \ \ \ \  \ (\gamma_{j,k}(u))_{j,k\leq N}\ \ \ \ \ \  \text{ with }\ \ \  \ \ \ u\in 
L^2((0,T),\R),
\]
 are skew-hermitian 
matrices. Thus, we can prove the solvability of \eqref{moment_problem_second} for $k<j$ and $j=k=1$. Now, we have 
$\big({x_{k,j}^{u_0}}\big)_{\underset{k\leq N }{j,k\in\N^*}}\in(h^3)^N$ and 
$\big({\gamma_{k,j}^{u_0}}\big)_{\underset{k\leq N}{j,k\in\N^*}}\in(h^3)^N$. Lemma 
\ref{perturbation_mixing_eigenspaces} yields that 
\[
\big({x_{k,j}^{u_0}}/ {B^{u_0}_{k,j}}\big)_{\underset{k\leq N}{j,k\in\N^*}}\in(\ell^2(\C))^N,\ \ \ \ \ \  \ \ 
\big({\gamma_{k,j}^{u_0}}/ {B^{u_0}_{k,j}}\big)_{\underset{k\leq N}{j,k\in\N^*}}\in(\ell^2(\C))^N.
\] Thanks to 
Lemma \ref{resonances_perturbation}, for $I^N$ defined in \eqref{I}, there exists 
\begin{equation*}\begin{split}
\Gi'&:=\sup_{A\subset 
I^N}\Big(\inf_{\underset{(j,k)\neq(n,m)}{(j,k),(n,m)\in I^N\setminus 
A}}|\lambda_j^{u_0}-\lambda_k^{u_0}-\lambda_n^{u_0}+\lambda_m^{u_0}|\Big)\\
&\geq 
\Gi:=\inf_{\underset{(j,k)\neq(n,m)}{(j,k),(n,m)\in I^N\setminus 
A}}|\lambda_j^{u_0}-\lambda_k^{u_0}-\lambda_n^{u_0}+\lambda_m^{u_0}|>0
\end{split}\end{equation*}
	where $A$ runs over the finite subsets of $I^N$ (we refer to the second point of the proof for further details 
on $\Gi'$). The solvability of the moment problem \eqref{moment_problem_second} is guaranteed from Corollary 
\ref{solvability_moment_problem} for 
$T>\frac{2\pi}{\Gi'}$ by considering the sequence of numbers obtained by reordering
\[
\big(\lambda_j^{u_0}-\lambda_k^{u_0}\big)_{\underset{k<j \text{ or }j=k=1}{{j,k\in\N^*},\ {k\leq N}}}\ .
\]	
Indeed, 
$x_{1,1}^{u_0}=0$ and Remark \ref{remark_general_perturbation} ensures that 
$\lambda_j^{u_0}-\lambda_k^{u_0}\neq\lambda_l^{u_0}-\lambda_m^{u_0}$ for every $(j,k),(l,m)\in I^N$ (see 
\eqref{I}) such that $(j,k)\neq(n,m)$. In conclusion, the solvability of the moment problem implies the 
surjectivity of $\gamma^{u_0}$ and the inverse mapping theorem ensures the surjectivity of $\alpha^{u_0}$ in 
$Q^N_\epsilon$ for $T>0$ large and suitable $\epsilon$. The proof is achieved as discussed in Remark 
\ref{control_projection_equivalent}.

	\smallskip
	\needspace{3\baselineskip}
	
	\noindent
	{\bf 2)} We show that the controllability ensured in {\bf 1)} is valid for every positive time $T>0$ by 
proving that $\Gi'=+\infty$. Let
\[
A^M:=\{(j,n)\in (\N^*)^2|\ j,n\geq M;\ j\neq n\},\ \ \ \  \ M\in\N^*.
\]
Thanks to the identity \eqref{decomposition_perturbed_eigenvalue} 
from the proof of Lemma \ref{perturb_eigenvalues}, for $|u_0|$ small enough and for 
$j>n$, we have
	\begin{eqnarray}
& & \lambda_N^{u_0}\leq \lambda_N+O(|u_0|)\label{perturbations},\\
& & \lambda_j^{u_0}-\lambda_n^{u_0}\geq \lambda_j-\lambda_n-O(|u_0|)\geq 
\pi^2(2n+1)-O(|u_0|).\label{perturbations1}
\end{eqnarray}
Hence, for every $K\in\R$, there exists $M_K>0$ large enough such that $$\inf_{(j,n)\in 
A^{M_K}}|\lambda_j^{u_0}-\lambda_n^{u_0}|>K.$$ Now,
\begin{equation*}\begin{split}
\Gi'&\geq \sup_{A\subset 
I^N}\Big(\inf_{\underset{(j,k)\neq(n,m)}{(j,k),(n,m)\in I^N\setminus 
A}}|\lambda_j^{u_0}-\lambda_n^{u_0}|-|\lambda_k^{u_0}+\lambda_m^{u_0}|\Big)\\
&\geq 
\sup_{M\in\N^*}\big(\inf_{(j,n)\in A^M}|\lambda_j^{u_0}-\lambda_n^{u_0}|-2\lambda_N^{u_0}\big)>0
\end{split}\end{equation*}
 where $A$ are 
the subsets of $I^N$ defining $\Gi'.$ In conclusion, for $|u_0|$ small enough, the relations 
\eqref{perturbations}-\eqref{perturbations1} yield	\begin{equation*}\begin{split}
	\Gi' &\geq \lim_{M\rightarrow\infty}\big(\inf_{(j,n)\in 
A^M}|\lambda_j-\lambda_n|-2\lambda_N-O(|u_0|)\big)\\
&\geq 
\lim_{M\rightarrow\infty}2M+1-2N^2\pi^2-O(|u_0|)=+\infty.	\end{split}\end{equation*}
	Finally, $\Gi'=+\infty$ and then the local exact controllability proved in the first point of 
the proof holds for every positive time since the result is valid for every $T>\frac{2\pi}{\Gi'}.$ 
\qedhere
\end{proof}

\section[Simultaneous global controllability]{Simultaneous global exact controllability in 
projection}\label{lollol}
\subsection{Preliminaries}
The common approach adopted in order to prove global exact controllability results consists in gathering the 
global 
approximate controllability and the local exact controllability. Nevertheless, this strategy can not be used to 
prove the controllability in projection as the propagator $\G_T^{u}$ does not preserve the space 
$\pi_N(\Psi)H^3_{(0)}$ for any $\Psi:=(\psi_j)_{j\in\N^*}\subset H^3_{(0)}$. For instance, let 
$\Psi=(\psi_j)_{j\in\N^*}$ be an orthonormal system and $\psi^1,\psi^2\in H^3_{(0)}$ be unitarily equivalent. 
Even though there exist $T_1,T_2>0,$ $u_1\in 
L^2((0,T_1),\R)$ and $u_2\in L^2((0,T_1),\R)$ such that
\[
\pi_N(\Psi)\G_{T_1}^{u_1}\psi_1=\pi_N(\Psi)\G_{T_2}^{u_2}\psi_2,
\]
it is not guaranteed the existence $T>0$ and a control $u\in 
L^2((0,T),\R)$ such that 
\[
\pi_N(\Psi)\G_{T}^{u}\psi_1=\pi_N(\Psi)\psi_2.
\]

To this purpose, we adopt an alternative strategy based on the result presented in the following subsection. 
There, 
we prove that the controllability in projection of infinite bilinear Schr\"odinger equations is equivalent to the 
controllability (without projecting) of a finite number of them. Hence, we ensure the simultaneous global exact 
controllability for $N\in\N^*$ \eqref{main_equation} in $(H^3_{(0)})^N$. In such space, we can concatenate and 
reverse 
dynamics as it is preserved by the dynamics. The result leads to Theorem \ref{main_theorem}.

\subsection{Equivalence between controllability of finite bilinear Schr\"odinger equations and infinitely many 
equations in projection}\label{provaequivalenza}

\begin{proposition}\label{equivalenza}
The two following assertions are equivalent with $N\in\N^*$.

 \noindent
 {\bf (1)} Let $(\psi_j^1)_{j\in\N^*}$ and $(\psi_j^2)_{j\in\N^*}\subset H^3_{(0)}$ be a couple of complete 
orthonormal systems of $\Hi$. Let $\widehat\G$ be the unitary operator such that 
$(\widehat\G\psi_j^2)_{j\in\N^*}=(\psi_j^1)_{j\in\N^*}$. For any $\Psi:=(\psi_j)_{j\leq N}\subset H^3_{(0)}$ 
orthonormal system of $\Hi$ such that $(\widehat\G\psi_j)_{j\leq N}\subset H^3_{(0)}$, there exist $T>0$ and 
$u\in L^2((0,T),\R)$ such that
	\begin{equation*}\begin{split}
	\la\psi_k,\G^u_T\psi_j^1\ra_{L^2}&=\la\psi_k,\psi_j^2\ra_{L^2},\ \ \ \ \ \forall j,k\in\N^*,\ k\leq N.\\
	\end{split}
	\end{equation*}
 In other words, the following identities are satisfied (with $\pi_N(\Psi)$ defined in \eqref{projector1}):
	\begin{equation*}\begin{split}
\pi_N(\Psi)\G_T^u\psi_j^1=\pi_N(\Psi)\psi^2_j,\ \ \ \ \ \ \ \ \ &\forall j\in\N^*.\\
	\end{split}
	\end{equation*}

 \noindent
 {\bf (2)}  Let $(\psi_j^1)_{j\leq N}$ and $(\psi_j^2)_{j\leq N}\subset H^3_{(0)}$ be a couple of orthonormal 
systems in $\Hi$. There exist $T>0$ and $u\in L^2((0,T),\R)$ such that
\begin{equation*}\begin{split}
\G_T^u\psi_j^1=\psi^2_j,\ \ \ \ \ \ \ \ \ &\forall j\leq N.\\
	\end{split}
	\end{equation*}
\end{proposition}
\begin{proof}
{\bf (2)\ $\Longrightarrow$\ (1) } Let $\Psi^3:=(\psi_j^3)_{j\in\N^*}\in H^3_{(0)}$ be an orthonormal system. We 
consider $(\psi_j^1)_{j\in\N^*},$ $(\psi_j^2)_{j\in\N^*}\subset  H^3_{(0)}$ complete orthonormal systems. Let 
$\widehat\G\in\ U(\Hi)$ be such that $\widehat\G\psi_j^2=\psi_j^1$ and $\widehat\G\psi_k^3\in H^3_{(0)}$ for every 
$k\leq N$. We notice that the controllability stated in the point {\bf (2)} of Theorem \ref{equivalenza} is also 
valid for the reversed dynamics discussed in Section \ref{time_revers}. Hence, there exist $T>0$ and $u\in 
L^2((0,T),\R)$ such that 
\[
\widetilde\G_{T}^{u}\psi_k^3=\widehat\G\psi_k^3,\ \ \ \ \forall k\leq N.
\]
Thus,
\[
\la\widetilde\G_{T}^{ u}\psi_k^3,\psi^1_j\ra_{L^2} 
=\la\widehat\G\psi_k^3,\psi^1_j\ra_{L^2},\ \ \ \ \  \ \forall j,k\in\N^*,\ k\leq N.
\]
Let $\widetilde u$ be introduced in Section \ref{time_revers}. The claim is proved since, for every $j,k\in\N^*$ 
with $ 
k\leq N$,
	\begin{equation*}\la \G_{T}^{\widetilde u}\psi_j^1, \psi_k^3\ra_{L^2}=\la 
\psi^1_j,\widetilde\G_{T}^{\widetilde u}\psi_k^3\ra_{L^2} 
=\la\psi^1_j,\widehat\G\psi_k^3\ra_{L^2}=\la\psi^2_j,\psi_k^3\ra_{L^2}.\end{equation*}

\smallskip

\noindent
{\bf (1)\ $\Longrightarrow$\ (2)\ } Let $(\psi_j^1)_{j\leq N},$ $(\psi_j^2)_{j\leq N}\subset  H^3_{(0)}$ be two 
orthonormal systems of $\Hi$. We complete them by defining $(\psi_j^1)_{j\in\N^*},$ $(\psi_j^2)_{j\in\N^*}\subset  
H^3_{(0)}$ two complete orthonormal systems of $\Hi$. Now, thanks to the point {\bf (1)}, there 
exist $T>0$ and $u\in L^2((0,T),\R)$ such that
\begin{equation*}\begin{split}
\pi_N(\Psi^2)\G^u_T\psi_j^1=\pi_N(\Psi^2)\ \psi_j^2,\ \ \ \ \ \ \ \ \ \ \ \ \ \  \forall j\in\N^*.\\
\end{split}\end{equation*}
As $\Psi^2$ is composed by orthogonal elements and $\G_T^u$ is unitary, the claim is proved since
\begin{equation*}\begin{split}\begin{cases}
	\G^u_T\psi_j^1=\psi_j^2,\ \ \ \ \ \ \ \ \ \ \ \ \ \ \ \ \ \ \ \ \ \ \ \ \ \ \ \ \ \ \ \ & \forall j\leq N,\\
	\pi_{N}(\Psi^2)\ \G^u_T\psi_j^1=0, & \forall j>N.\qedhere
	\end{cases}
	\end{split}
	\end{equation*}
\end{proof}

\begin{remark}\label{compatibility}
 The previous proof contains the reason why we need to impose a $H^3_{(0)}-$compatibility condition such as 
\eqref{compatibilitycondition1} in order to obtain the controllabilty in projection of infinitely many 
\eqref{main_equation}.  In particular, let $T>0$, $u\in L^2((0,T),\R)$, $\widehat\Gamma\in U(\Hi)$, 
$(\psi_j)_{j\in\N^*}\subset H^3_{(0)}$ and $(\psi^1_j)_{j\in\N^*}\subset H^3_{(0)}$ be a complete orthonormal 
system of $\Hi$. We know that if for every $j,k\in\N^*$ and $k\leq N$, we have
 \[
\la\G_T^u\psi^1_j,\psi_k\ra_{L^2}=\la\widehat\G\psi^1_j,\psi_k\ra_{L^2},\ \ \ \ \ \Longrightarrow\ \ \ \ \ \  
\la\psi^1_j,\widetilde \G_T^{\widetilde u}\psi_k\ra_{L^2}=\la\psi^1_j,\widehat\G\psi_k\ra_{L^2}.
\]
 The last relation is equivalent to $\widetilde \G_T^{\widetilde u}\psi_k=\widehat\G\psi_k$ for every $k\leq N.$ 
Now, $\widetilde \G_T^{\widetilde u}$ is the propagator of the reversed dynamics introduced in the previous 
section 
and it preserves $H^3_{(0)}$. This fact tells that the controllability in projection can be ensured only when a 
$H^3_{(0)}-$compatibility condition such as \eqref{compatibilitycondition1} is guaranteed. Namely, when 
$\widehat\Gamma$ and $(\psi_j)_{j\in\N^*}$ are such that $\widehat\G\psi_k\in H^3_{(0)}$ for 
every $k\leq N.$
 \end{remark}

\subsection{Simultaneous approximate controllability}\label{approximate}
In this section, we prove the simultaneous global approximate controllability for finite number of 
\eqref{main_equation}.
\begin{definition}
	The problems \eqref{main_equation} are said to be simultaneously globally approximately controllable in 
$H_{(0)}^{3}$ when, for every $N\in\N^*$, $\psi_1,....,\psi_N\in H^{3}_{(0)}$, $\widehat\G\in U(\Hi)$ such that 
$\widehat\G\psi_1,....,\widehat\G\psi_N\in H^{3}_{(0)}$ and $\epsilon>0$, there exist $T>0$ and $u\in 
L^2((0,T),\R)$ such that $\|\G^u_T\psi_k-\widehat\G\psi_k\|_{(3)}<\epsilon$ for every $1\leq k\leq N$.
\end{definition}
\begin{theorem}\label{theorem_approx}
	Let $B$ satisfy Assumptions I. The problems \eqref{main_equation} are simultaneously globally approximately 
controllable in $H^{3}_{(0)}$.
\end{theorem}
\begin{proof}

In the point {\bf 1)\,\,}of the proof, we suppose that $(A,B)$ admits a non-degenerate chain of connectedness (see 
\cite[Definition\ 3]{nabile}). We treat the general case in the point {\bf 2)\,\,}of the proof.

\smallskip

\needspace{3\baselineskip}

\noindent
{\bf 1) Preliminaries.} Let $\pi_m$ be the orthogonal projector $\pi_m:\Hi\rightarrow \Hi_m:=span\{\phi_j\ 
:\ j\leq m\}$ for every $m\in\N^*.$
Up to reordering of $(\phi_k)_{k\in\N^*}$, the couples $(\pi_{m}A\pi_{m},\pi_{m}B\pi_{m})$ for $m\in\N^*$ admit 
non-degenerate chains of connectedness in $\Hi_{m}$. Let $\|\cdot\|_{BV(T)}=\|\cdot\|_{BV((0,T),\R)}$ and 
$\iii\cdot\iii_{(s)}:=\iii\cdot\iii_{L(H^s_{(0)},H^s_{(0)})}$ for $s>0.$
Thanks to the validity of Assumptions I, we have $B:H^{2}_{(0)}\rightarrow H^{2}_{(0)}$. 
Let us denote  
\[
SU(\Hi_{m})=\big\{\Gamma\in U(\Hi_{m})\ :\ (\la\phi_j,\Gamma\phi_k\ra_{L^2})_{j,k\leq m}\in 
SU(m)\big\}.
\]

	\smallskip
\noindent
\underline{\bf Claim.} For every $\epsilon>0$, there exist $N_1\in\N^*$ with $N_1\geq N$ and 
$\widetilde\G_{N_1}\in U(\Hi)$ such that $\pi_{N_1}(\Phi)\widetilde\G_{N_1}\pi_{N_1}(\Phi)\in SU(\Hi_{N_1})$ and
\begin{equation}\label{finite_dimensional_unitary}\|\widetilde\G_{N_1}\phi_j-\widehat\Gamma\phi_j\|_{L^2}<\epsilon,
\ \ \ \ \ \forall 
j\leq N.\end{equation}

	\smallskip

	Let $N_1\in\N^*$ be such that $N_1\geq N$. We apply the orthonormalizing Gram-Schmidt process to 
$(\pi_{N_1}(\Phi)\widehat\G\phi_j)_{j\leq N}$ and we define the sequence $(\widetilde\phi_j)_{j\leq N}$ that we 
complete in $(\widetilde\phi_j)_{j\leq N_1}$, an orthonormal basis of $\Hi_{N_1}$. We complete again such 
sequence in an orthonormal basis of $\Hi$ that we call $(\widetilde\phi_j)_{j\in\N^*}$. The operator 
$\widetilde\G_{N_1}$ is 
the unitary map such that $\widetilde\G_{N_1}\phi_j=\widetilde\phi_j,$ for every $j\in\N^*.$ In conclusion, we 
consider $N_1$ large enough such that the statement is verified.
	
\smallskip

	\needspace{3\baselineskip}
	
	\noindent
	{\bf Finite dimensional controllability.} Let $T_{ad}$ be the set of $(j,k)\in\{1,...,N_1\}^2$ such that 
$B_{j,k}:=\la\phi_j,B\phi_k\ra_{L^2}\neq 0$ and $|\lambda_j-\lambda_k|=|\lambda_m-\lambda_l|$ with $m,l\in\N^*$ 
implies $\{j,k\}= \{m,l\}$ for $B_{m,l}=0$.
For every $(j,k)\in\{1,...,N_1\}^2$ and $\theta\in[0,2\pi)$, we define $E_{j,k}^{\theta}$ the $N_1\times N_1$ 
matrix with elements $$(E_{j,k}^\theta)_{j,k}=e^{i\theta},\ \ \ \ 
\ (E_{j,k}^\theta)_{k,j}=-e^{-i\theta},\ \ \ \ \ \ (E_{j,k}^\theta)_{l,m}=0,$$ for 
$(l,m)\in\{1,...,N_1\}^2\setminus\{(j,k),(k,j)\}.$ Let 
$$E_{ad}=\big\{E_{j,k}^{\theta}\ :\ (j,k)\in T_{ad},\ \theta\in[0,2\pi)\big\}$$ and $Lie(E_{ad})$. 
 Fixed 
$v$ a piecewise constant control taking value in $E_{ad}$ and 
$\tau>0$, we introduce the control system on $SU({N_1})$
\begin{equation}\label{control_finite}\begin{split}\begin{cases}
\dot{x}(t)=x(t)v(t),\ \ \ \ \ \ t\in(0,\tau),\\
x(0)=Id_{SU({N_1})}.\\
\end{cases}\end{split}\end{equation}
\smallskip

\noindent
\underline{\bf Claim.} \eqref{control_finite} is controllable, {\it i.e.} for $R\in SU({N_1})$, there exist 
$p\in\N^*$, $M_1,...,M_p\in E_{ad}$, $\alpha_1,...,\alpha_p\in\R^+$ such that $R=e^{\alpha_1 M_1}\circ...\circ 
e^{\alpha_p M_p}.$
\smallskip

For every $(j,k)\in\{1,...,N_1\}^2$, we define the $N_1\times N_1$ matrices $R_{j,k}$, $C_{j,k}$ and $D_{j}$ as 
follows. For $(l,m)\in\{1,...,N_1\}^2\setminus\{(j,k),(k,j)\},$ we have 
\[
 (R_{j,k})_{l,m}=0,\ \ \ \  \ \ (R_{j,k})_{j,k}=-(R_{j,k})_{k,j}=1,
 \]
\[
(C_{j,k})_{l,m}=0,\ \ \ \ \ \ (C_{j,k})_{j,k}=(C_{j,k})_{k,j}=i.
\]
Moreover, 
for $(l,m)\in\{1,...,N_1\}^2\setminus\{(1,1),(j,j)\},$ $$(D_{j})_{l,m}=0,\ \ \ \ \ 
\ (D_{j})_{1,1}=-(D_{j})_{j,j}=i.$$
We denote by $su({N_1})$ the Lie algebra of 
$SU({N_1})$ and we consider its basis
\[
{\bf e}:=\{R_{j,k}\}_{j,k\leq N_1}\cup\{C_{j,k}\}_{j,k\leq 
N_1}\cup\{D_{j}\}_{j\leq N_1}.
\]
Thanks to \cite[Theorem\ 6.1]{sac}, the controllability of (\ref{control_finite}) 
is equivalent to prove that $Lie(E_{ad})\supseteq su({N_1})$. The claim is valid as it is possible to obtain 
the matrices  $R_{j,k}$, $C_{j,k}$ and $D_j$ for every $j,k\leq N_1$ by iterated Lie brackets of elements in 
$E_{ad}$.

\smallskip

	\needspace{3\baselineskip}
	
	\noindent
	{\bf Finite dimensional estimates.} The previous claim and the fact that the matrix 
$(\la\phi_j,\widetilde\G_{N_1}\phi_k\ra_{L^2})_{j,k\leq N_1}\in SU({N_1})$ ensure the existence of $p\in\N^*$, 
$M_1,...,M_p\in E_{ad}$ and $\alpha_1,...,\alpha_p\in\R^+$ such that
	\begin{equation}\label{decomposition_rotations}(\la\phi_j,\widetilde\G_{N_1}\phi_k\ra_{L^2})_{j,k\leq 
N_1}=e^{\alpha_1 
M_1}\circ...\circ e^{\alpha_p M_p}.\end{equation}
For every $l\leq p$, we call
$\widehat\Gamma_l$ the operator in $SU(\Hi_{N_1})$ such that $(\la\phi_j,\widehat\Gamma_l\phi_k\ra)_{j,k\leq 
N_1}=e^{\alpha_l M_l}$. The identity \eqref{decomposition_rotations} yields
\begin{equation}\label{decomposition_unitary_rotations}\pi_{N_1}(\Phi)\widetilde\G_{N_1}\pi_{N_1}
(\Phi)=\widehat\Gamma_1\circ...\circ\widehat\Gamma_p.\end{equation}
\smallskip

\noindent
\underline{\bf Claim.} For every $l\leq p$ and $\widehat\Gamma_l$ from 
\eqref{decomposition_unitary_rotations}, there 
exist $(T_n^l)_{l\in\N^*}\subset\R^+$ and $(u_n^l)_{n\in\N^*}$ such that $u_n^l:(0,T_n^l)\rightarrow\R$ for 
every $n\in\N^*$ and
\begin{equation}\label{approximate_control_roation}\lim_{n\rightarrow\infty}\|\G_{T_n^l}^{u_n^l}
\phi_k-\widehat\Gamma_l\phi_k\|_{L^2}=0,
\ \ \ \  \ \ \forall k\leq N_1, \end{equation}
		\begin{equation}\label{bound_uniform_approx_control}\begin{split}\sup_{n\in\N^*}\|u_n^l&
		\|_{BV(T_n)}<\infty,  \ \ \  \ \ \ \ \ \sup_{n\in\N^*}\|u_n^l\|_{L^\infty((0,T_n),\R)}<\infty,\\ 
		&\sup_{n\in\N^*} T_n\|u_n^l\|_{L^\infty((0,T_n),\R)}<\infty.\end{split}\end{equation}

		\smallskip

		We consider the results developed in \cite[Section\ 3.1\ \&\ Section\ 3.2]{chambrion2} by Chambrion and 
leading to \cite[Proposition\ 6]{chambrion2} since $(A,B)$ admits a non-degenerate chain of connectedness 
(defined in \cite[Definition\ 3]{nabile}). Each $\widehat\Gamma_l$ corresponds to a rotation in a two dimensional 
space for 
every $l\in\{1,...,p\}$. This work allows to explicit $(T_n^l)_{l\in\N^*}\subset\R^+$ and 
$(u_n^l)_{n\in\N^*}$ satisfying \eqref{bound_uniform_approx_control} such that each 
$u_n^l:(0,T_n^l)\rightarrow\R$ 
and 
\[
 \lim_{n\rightarrow\infty}\|\pi_{N_1}(\Phi)\G_{T_n^l}^{u_n^l}
\phi_k-\widehat\Gamma_l\phi_k\|_{L^2}=0,\ \ \ \ \ \ \forall k\leq N_1.
\]
As $\widehat\Gamma_l\in SU(\Hi_{N_1})$,  we have 
$\lim_{n\rightarrow\infty}\|\G_{T_n^l}^{u_n^l}\phi_k-\widehat\Gamma_l\phi_k\|_{L^2}=0$ for $k\leq N_1.$

\smallskip

	\needspace{3\baselineskip}
	
	\noindent
	{\bf Infinite dimensional estimates.} 
 
\smallskip
 \noindent
 \underline{\bf Claim.} There exist $K_1,K_2,K_3>0$ such that for every $\epsilon>0$, there exist $T>0$ and 
$u\in L^2((0,T),\R)$ such that $\|\G_{T}^{u}\phi_k-\widehat\G\phi_k\|_{L^2}<\epsilon$ for every $k\leq N_1$ and
\[
\|u\|_{BV(T)}\leq K_1,  \ \  \ \ \ \ \ \|u\|_{L^\infty((0,T),\R)}\leq K_2,\ \  
\ \ \ \ \ T\|u\|_{L^\infty((0,T),\R)}\leq K_3.
\]
\smallskip

	Let us assume that {\bf 1) (c)} be valid with $p=2$. Nevertheless, the following result is valid for any 
$p\in\N^*$. By definition of $\widehat\Gamma_1\in SU(\Hi_{N_1})$, for every $k\leq N_1$, there exist $ 
l_k\leq N_1$ and $\alpha_{l_k}\in\C$ with $|\alpha_{l_k}|=1$ such that 
$\widehat\Gamma_1\phi_k=\alpha_{l_k}\phi_{l_k}$. Thanks to \eqref{approximate_control_roation}, for $n\in\N^*$ 
sufficiently large,
	\begin{equation*}\begin{split}\| \G_{T_n^2}^{u_n^2}\G_{T_n^1}^{u_n^1}\phi_k-\widehat\Gamma_2 
\widehat\Gamma_1\phi_k\|_{L^2}
\leq &\iii 
\G_{T_n^2}^{u_n^2}\iii\|\G_{T_n^1}^{u_n^1}\phi_k-\widehat\Gamma_1\phi_k\|_{L^2}\\
&+\|\alpha_{l_k}\G_{T_n^2}^{u_n^2}
\phi_{l_k}-\alpha_{l_k}\widehat\Gamma_2\phi_{l_k}\|_{L^2}<\epsilon,\ \ \ \ \ \ \ \forall k\leq 
N_1.\end{split}\end{equation*}
The identity 
\eqref{decomposition_unitary_rotations} leads to the existence of $K_1,K_2,K_3>0$ such that for every 
$\epsilon>0$, there exist $T>0$ and 
$u\in L^2((0,T),\R)$ such that $\|\G_{T}^{u}\phi_k-\widetilde\G_{N_1}\phi_{k}\|_{L^2}<\epsilon$ for every 
$k\leq 
N_1$ and 
\[
\|u\|_{BV(T)}\leq K_1,   \ \ \ \ \ \|u\|_{L^\infty((0,T),\R)}\leq K_2,\  \ \ \ \ T\|u\|_{L^\infty((0,T),\R)}\leq 
K_3.
\]
	The relation \eqref{finite_dimensional_unitary} and the triangular inequality achieve the claim.
	
	\smallskip

	\noindent
	{\bf Global approximate controllability with respect to the $L^2$-norm.} Let us recall that $(\psi_j)_{j\leq 
N}\subset H^3_{(0)}$ and $\widehat\G\in U(\Hi)$ satisfies $(\widehat\G\psi_j)_{j\leq N}\subset H^3_{(0)}$.
	
		\smallskip

		\noindent
		\underline{\bf Claim.} There exist $K_1,K_2,K_3>0$ such that for every $\epsilon>0$, there exist $T>0$ and 
$u\in L^2((0,T),\R)$ such that $\|\G_{T}^{u}\psi_k-\widehat\G\psi_k\|_{L^2}<\epsilon$ for every $k\leq N$ and
		\begin{align}\label{upper_bound_approx_control1}\|u\|_{BV(T)}\leq K_1,  \ \  \ \ \ \ \ 
\|u\|_{L^\infty((0,T),\R)}\leq K_2,\ \  \ \ \ \ \ T\|u\|_{L^\infty((0,T),\R)}\leq K_3.\end{align}

    We assume that $\|\psi_j\|_{L^2}=1$ for every $j\in\N^*$, but the same proof is also valid for the generic 
case.
	From the previous claim, there exist two controls respectively steering $(\phi_j)_{j\leq N}$ close to 
$(\psi_j)_{j\leq N}$ and $(\phi_j)_{j\leq N}$ close to $(\widehat\G\psi_j)_{j\leq N}$ thanks to the fact that 
$N_1\geq N$. Vice versa, thanks to the time reversibility (see Section \ref{time_revers}), there exists a control 
steering $(\psi_j)_{j\leq N}$ close to $(\phi_j)_{j\leq N}$. In other words, there exist $T_1, T_2>0$, $u_1\in 
L^2((0,T_1),\R)$ and $u_2\in L^2((0,T_2),\R)$ such that
\[
\|\G_{T_1}^{u_1}\psi_j-\phi_j\|_{L^2}<\frac{\epsilon}{2},\ \ \ \  \ \ \ \ \ 
\|\G_{T_2}^{u_2}\phi_j-\widehat\G\psi_j\|_{L^2}<\frac{\epsilon}{2},\ \ \ \ \ \ \forall j\leq N.
\]
The chosen controls $u_1$ and $u_2$ satisfy \eqref{upper_bound_approx_control1}. The claim is proven as
\[
\|\G_{T_2}^{u_2}\G_{T_1}^{u_1}\psi_j-\widehat\G\psi_j\|_{L^2}\leq 
\|\G_{T_2}^{u_2}\G_{T_1}^{u_1}\psi_j-\G_{T_2}^{u_2}\phi_j\|_{L^2}+\|\G_{T_2}^{u_2}\phi_j-\widehat\G\psi_j\|_{L^2}
<{\epsilon},\ \ \ \ \ \forall j\leq N.
\]
	
\smallskip

	\noindent
	{\bf Global approximate controllability with respect to the $H^3_{(0)}$-norm.} 
\smallskip

\noindent
\underline{\bf Claim.} There exist $T>0$ and $u\in L^2((0,T),\R)$ such that 
$\|\G_{T}^{u}\psi_k-\widehat\G\psi_k\|_{(3)}<\epsilon$ for every $k\leq N$.

\smallskip

	We consider the propagation of regularity developed by Kato in \cite{kato1}. We notice that $i(A+u(t)B-ic)$ 
is maximal dissipative in $H^2_{(0)}$ for suitable $c>0$. Let $\lambda>c$ and $\widehat 
H^4_{(0)}:=D(A(i\lambda-A))\equiv  H^4_{(0)}$. We know that $B:\widehat H^4_{(0)}\subset  H^2_{(0)}\rightarrow 
H^2_{(0)}$ and the arguments of Remark \ref{osss_closed} imply that $B\in L(\widehat H^4_{(0)},H^2_{(0)})$. For 
$T>0$ and $u\in BV((0,T),\R)$, we have $\iii u(t)B(i\lambda-A)^{-1}\iii_{(2)}< 1$ and 
	\begin{equation*}\begin{split}
	M&:=\sup_{t\in [0,T]}\iii(i\lambda-A-u(t)B)^{-1}\iii_{L(H^2_{(0)},\widehat H^4_{(0)})}\\
	&\leq\sup_{t\in 
[0,T]}\sum_{l=1}^{+\infty}\iii(u(t)B(i\lambda-A)^{-1})^l\iii_{(2)}<+\infty.\\
	\end{split}\end{equation*}
We know $\|k+f(\cdot)\|_{BV((0,T),\R)}=\|f\|_{BV((0,T),\R)}$ for $f\in BV((0,T),\R)$ and $k\in\R$. Equivalently, 
	\begin{equation*}\begin{split}
N&:=\iii i\lambda-A-u(\cdot)B\iii_{BV\big([0,T],L(\widehat H^4_{(0)},H^2_{(0)})\big)}\\
&=\|u\|_{BV(T)} \iii 
B\iii_{L(\widehat H^4_{(0)},H^2_{(0)})}<+\infty.
	\end{split}\end{equation*}
We call $U_t^{u}$ the propagator generated by $A+uB-ic$ such that $U_t^u\psi=e^{-ct}\G_t^u\psi$. Thanks to 
\cite[Section\ 3.10]{kato1}, for every $\psi\in H^4_{(0)}$, it follows 
\begin{equation*}\begin{split}\|(A+u(T)B-i\lambda) U_t^u \psi\|_{(2)}\leq Me^{MN}\|(A-i\lambda) \psi\|_{(2)}	
\end{split}\end{equation*}
which implies, for $C_1:=\iii A(A+u(T)B-i\lambda)^{-1}\iii_{(2)}<\infty$,
\begin{equation*}\begin{split}\|\G_{T}^{u} \psi\|_{(4)}&\leq C_1 
Me^{MN+cT}\|\psi\|_{(4)}.\end{split}\end{equation*}
For every $T>0$, $u\in BV((0,T),\R)$ and $\psi\in H^{4}_{(0)}$, there exists $C(K)>0$ depending on 
$K=\big(\|u\|_{BV(T)},\|u\|_{L^\infty((0,T),\R)},T\|u\|_{L^\infty((0,T),\R)}\big)$ such that $\|\G_{T}^{u} 
\psi\|_{(4)}\leq  C(K)\|\psi\|_{(4)}.$ When \eqref{bound_uniform_approx_control} is verified, there exists $C>0$ 
such that, for every $n\in\N^*$,
\begin{equation}\label{bound_propagatore_H4}\iii\G_{T_n^l}^{u_n^l}\iii_{(4)}\leq C.\end{equation}
For every $\psi\in H^4_{(0)}$, from the Cauchy-Schwarz inequality, we have $\|A\psi\|_{L^2}^2\leq\la 
A^2\psi,\psi\ra_{L^2}
\leq\|A^2\psi\|_{L^2}\|\psi\|_{L^2}$ and  $\|A^\frac{3}{2}\psi\|_{L^2}^4\leq\big(\la A^2\psi, A\psi\ra_{L^2}\big)^2
\leq\|A^2\psi\|_{L^2}^2\|A\psi\|_{L^2}^2,$ which imply
\begin{equation}\label{bound_propagatore_H3}\|\psi\|_{(3)}^8\leq\|\psi\|_{L^2}^2\|\psi\|_{(4)}^6.\end{equation}
	In conclusion, the claim of the global approximate controllability with respect to the $L^2$-norm and the 
relations \eqref{bound_propagatore_H4}-\eqref{bound_propagatore_H3} 
ensure the claim.

\smallskip

	\noindent
	{\bf 2) Conclusion.} Assume that $(A,B)$ does not admit a non-degenerate chain of connectedness. We decompose 
\[
A+u(\cdot)B=(A+u_0B)+u_1(\cdot)B,\ \ \ \ \ \ \  \ \ \ \ \ \  \ u_0\in \R,\ \ \ \ u_1\in L^2((0,T),\R).
\] 
We 
notice that, if $(A,B)$ satisfies Assumptions I, then Remark \ref{perturbated_norm_equivalence_remark} and Remark 
\ref{remark_general_perturbation} are 
valid. We consider $u_0$ belonging to the neighborhoods provided by such remarks and we 
denote by $(\phi_k^{u_0})_{k\in\N^*}$ a complete orthonormal system of $\Hi$ made by eigenfunctions of $A+u_0B$.
Thanks to the first point of Remark \ref{remark_general_perturbation}, the couple $(A+u_0B,B)$ admits a 
non-degenerate chain of 
connectedness. The 
step {\bf 1)} of the proof can be repeated by considering the sequence $(\phi_k^{u_0})_{k\in\N^*}$ instead of 
$(\phi_k)_{k\in\N^*}$ and the spaces $D(|A+u_0B|^\frac{3}{2})$ in substitution of $H^{3}_{(0)}$. The claim is 
equivalently proved since, thanks to Remark \ref{perturbated_norm_equivalence_remark}, there exist $C_1,C_2>0$ 
such that 
\[
C_1\big\||A+u_0B|^\frac{3}{2}\psi\big\|\leq\|\psi\|_{(3)}\leq C_2\big\||A+u_0B|^\frac{3}{2}\psi\big\|,\ \ \ \  \ 
\ \ \ \forall \psi\in H^3_{(0)}.\qedhere
\]
\end{proof}

\subsection{Proof of Theorem \ref{main_theorem}}\label{proof_main_theorem}
In the current subsection, we provide the proof of Theorem \ref{main_theorem} which requires the following 
proposition.

\begin{proposition}\label{finite_simultaneous}
	Let $N\in\N^*$ and $B$ satisfy Assumptions I. For any $(\psi_k^1)_{k\leq N}$, $(\psi_k^2)_{k\leq N}\subset 
H^3_{(0)}$ orthonormal systems, there exist $T>0$, $u\in L^2((0,T),\R)$ and $(\theta_k)_{k\leq N}\subset \R$ such 
that 
\[
\G^u_T\psi_k^1=e^{i\theta_k}\psi_k^2,\ \ \ \ \ \ \forall k\leq N.
\]
\end{proposition}
\begin{proof}
	Let $N\in\N^*$ and let $u_0\in\R$ belong to the neighborhoods provided by Lemma 
\ref{perturbation_mixing_eigenspaces}, Lemma 
\ref{perturnated_norm_equivalence} and Remark \ref{remark_general_perturbation}. Let $\widetilde \alpha^{u_0}$ be 
the map with elements
	
\begin{equation*}\begin{split}&\begin{cases}
\frac{\overline{\widehat\alpha_{j,j}(u_1)}}{|\widehat\alpha_{j,j}(u_1)|
}\widehat\alpha_{ k,j}(u_1), \ \  \ \ \ \ \ \ \ \ \ \ \ \ \ \ \ \ &\forall j,k\in\N^*,\  j,k\leq N,\\
	\widehat\alpha_{k,j}(u_1),\ & \forall j,k\in\N^*,\  k>N,\ j\leq N.\\
	\end{cases}
	\end{split}
	\end{equation*}
	The proof of Proposition \ref{proposition_local_control} can be repeated in order to prove the local 
surjectivity of $\widetilde \alpha^{u_0}$ for every $T>0$, instead of $\alpha^{u_0}$ introduced in 
\eqref{defi_alfa}. The discussion from Remark \ref{control_projection_equivalent} implies that this result 
corresponds to the simultaneous local exact controllability up to phases of $N$ problems \eqref{main_equation} in 
the neighborhood
	\[
O^N_{\epsilon,T}:=\Big\{(\psi_j)_{j\leq N}\subset H^3_{(0)}\big|\ \la\psi_j,\psi_k\ra_{L^2}=\delta_{j,k};\ \ 
\sup_{j\leq N}\|\psi_j-\phi_j^{u_0}\|_{ (3)}^2<\epsilon\Big\}
\]
	with $\epsilon>0$. Hence, for any $(\psi_k)_{k\leq N}\in O^N_{\epsilon, T}$, there exist $u\in L^2((0,T),\R)$ 
and $(\theta_j)_{j\leq N}\subset\R$ such that 
\[
\G^{u}_{T}\phi_j^{u_0}=e^{i\theta_j}\psi_j,\ \ \ \ \ \  \ \forall 
j\leq N.
\]
Thanks to Theorem \ref{theorem_approx}, we have the following result. For any $(\psi^1_j)_{j\leq N}\subset 
H^3_{(0)}$ composed by orthonormal elements, there exist $T_1>0$ and $u_1\in L^2((0,T_1),\R)$ such that, for all 
$j\leq N$, 
\[
\|\G^{u_1}_{T_1}\psi^1_j-\phi_j^{u_0}\|_{(3)}<\frac{\epsilon}{N}\ \ \ \ \ \ \ \ \Longrightarrow\ \ 
 \  
\ \  \ \ \ (\G^{u_1}_{T_1}\psi^1_j)_{j\leq N}\in O^N_{\epsilon,T}.
\]
 The local controllability is also valid for 
the reversed dynamics (see Section \ref{time_revers}) and for every $T>0$, there exist $u\in L^2((0,T),\R)$ and 
$(\theta_j)_{j\leq N}\subset\R$ such that 
\[
(\G^{u_1}_{T_1}\psi^1_j)_{j\leq N}=(e^{i\theta_j}\widetilde 
\G_T^{u}\phi_j^{u_0})_{j\leq N}\ \ \  \ \ \Longrightarrow \ \ \ (e^{-i\theta_j}\G_T^{\widetilde 
u}\G^{u_1}_{T_1}\psi^1_j)_{j\leq N}=(\phi_j^{u_0})_{j\leq N}.
\]
Then, there exist $T_2>0$ and $u_2\in 
L^2((0,T_2),\R)$ such that $(e^{-i\theta_j}\G_{T_2}^{u_2}\psi^1_j)_{j\leq N}=(\phi_j^{u_0})_{j\leq N}.$ Now, 
the same property is valid for the reversed dynamics of \eqref{main_equation_reversed} and, for every 
$(\psi^2_j)_{j\leq 
N}\subset H^3_{(0)}$ composed by orthonormal elements, there exist $T_3>0$, $u_3\in L^2((0,T_3),\R)$ and 
$(\theta_j')_{j\leq N}\subset\R$ such that $(e^{-i\theta'_j}\widetilde\G_{T_3}^{u_3}\psi^2_j)_{j\leq 
N}=(\phi_j^{u_0})_{j\leq N}.$	In conclusion, for $\widetilde u_3(\cdot)=u_3(T_3-\cdot)$, the proof is achieved as
\[
(e^{-i(\theta_j-\theta_j')}\G_{T_3}^{\widetilde u_3}\G_{T_2}^{u_2}\psi^1_j)_{j\leq N}=(\psi^2_j)_{j\leq 
N}.\qedhere
\]
\end{proof}

\begin{proof}[Proof of Theorem \ref{main_theorem}]
	The claim is proved as the implication {\bf (2)\ $\Longrightarrow$ (1)}\\ in the proof of 
Theorem \ref{equivalenza} thanks to the validity of Proposition 
\ref{finite_simultaneous}.\qedhere
\end{proof}

\section{Global exact controllability in projection for density matrices}\label{density_matrices}

Let $\psi^1$, $\psi^2\in\Hi$. We define the rank one operator $|\psi^1\ra\la\psi^2|$ such that 
$|\psi^1\ra\la\psi^2|\psi=\psi^1\la\psi^2,\psi\ra_{L^2}$ for every $\psi\in\Hi.$ For any $\widehat\G\in U(\Hi)$, 
we have 
\[
\widehat\G|\psi^1\ra\la\psi^2|=|\widehat\G\psi^1\ra\la\psi^2|,\ \ \ \ \ \  \ \ 
|\psi^1\ra\la\psi^2|\widehat\G^*=|\psi^1\ra\la\widehat\G\psi^2|.
\]
Let $\Hi$ be an infinite dimensional Hilbert space. In quantum mechanics, any statistical ensemble can be 
described 
by a wave function (pure state) or by a density matrix (mixed state) which is a positive operator of trace $1$. 
For 
any density matrix $\rho$, there exists a sequence $(\psi_j)_{j\in \N^*}\subset\Hi$ such that
\begin{equation}\label{deco_density}
\begin{split}
\rho&=\sum_{j\in \N^*} l_j|\psi_j\ra\la \psi_j|,\ \ \ \ \sum_{j\in \N^*}l_j=1,\ \ \ \ l_j\geq 0,\ \ \forall j\in 
\N^*.
\end{split}
\end{equation}
The sequence $(\psi_j)_{j\in \N^*}$ is a set of eigenvectors of $\rho$ and $(l_j)_{j\in \N^*}$ are the 
corresponding eigenvalues.
If there exists $j_0\in\N^*$ such that $l_{j_0}=1$ and $l_j=0$ for each $j\neq j_0$, then the corresponding 
density 
matrix represents a pure state up to a phase. For this reason, the density matrices formalism is said to be an 
extension of the common formulation of the quantum mechanics in terms of wave function. We also notice that for 
any 
density matrix $\rho$ and a complete orthonormal system $(\psi_j)_{j\in\N^*}$ in $\Hi$, there exists a positive 
hermitian matrix $(\rho_{j,k})_{j,k\in\N^*}$ such that
\begin{align}\label{deco_density_orth_basis}
 \rho=\sum_{j,k\in\N^*}\rho_{j,k}|\psi_j\ra\la\psi_k|.
\end{align}
Now, for any other density matrix $\widetilde\rho$, there exists an orthonormal system 
$(\widetilde\psi_j)_{j\in\N^*},$ such that
\begin{align}\label{deco_density_orth_basis1}
 \widetilde\rho=\sum_{j,k\in\N^*}\rho_{j,k}|\widetilde\psi_j\ra\la\widetilde\psi_k|.
\end{align}
Let us consider $T>0$ and a time dependent self-adjoint operator $H(t)$ (called Hamiltonian) for $t\in(0,T)$. The 
dynamics of a general density matrix $\rho$ is described by the Von Neumann equation
\begin{equation}\label{Von_Neumann}\begin{split}
\begin{cases}
i\frac{d\rho}{dt}(t)=[H(t),\rho(t)],\  \ \ \ \ \ \ \  &\  \ \ \ \ \ \ \  \ \ \ \ \ t\in(0,T),\\
\rho(0)=\rho^0,  \ \ \ \ &([H,\rho]=H\rho-\rho H),\\
\end{cases}
\end{split}
\end{equation}
for $\rho^0$ the initial solution of the problem. The solution is $\rho(t)=U_t\rho(0)U_t^*$, where $U_t$ is the 
unitary propagator generated by $H(t)$. In the present work, we have $\Hi=L^2((0,1),\C)$, $H(t)=A+u(t)B$ and $U_t$ 
corresponds to $\G_t^u$. In this framework, the problem \eqref{Von_Neumann} is said to be globally exactly 
controllable if, for any couple of density matrices $\rho^1$ and $\rho^2$, there exist $T>0$ and $u\in 
L^2((0,T),\R)$ such that \[
\rho^2=\G_T^u\rho^1(\G_T^u)^*.
\]
Thanks to the decomposition \eqref{deco_density}, the controllability of \eqref{Von_Neumann} is equivalent (up 
to phases) to the simultaneous controllability of the infinite bilinear Schr\"odinger equations 
\eqref{main_equation}. This idea is behind the following theorem which follows from Corollary \ref{main_corollary}.

\begin{theorem}\label{main_theorem_density}
	Let $B$ satisfy Assumptions I. Let $\rho^1$ and $\rho^2$ be two density matrices with eigenfunctions in 
$H^3_{(0)}$ and $\widehat\G\in U(\Hi)$ be such that 
\[
\rho^1=\widehat\G \rho^2\widehat\G^*.
\]

	\noindent
	{\bf 1)} Let $\Psi:=(\psi_j)_{j\in \N^*}$ be an orthonormal system composed by the eigenfunctions of $\rho^2$. 
For any $N\in\N^*$, there exist $T>0$ and $u\in L^2((0,T),\R)$ such that
	\begin{equation*}\begin{split}
	\pi_{N}(\Psi)\ \G_{T}^{u}\rho^1(\G_{T}^{u})^*\ \pi_{N}(\Psi)= \pi_{N}(\Psi)\ \rho^2\ \pi_{N}(\Psi).\\
	\end{split}
	\end{equation*}

	\noindent
	{\bf 2)} Let $\Psi:=(\psi_j)_{j\leq N}\subset H^3_{(0)}$ be an orthonormal system such that 
$(\widehat\G\psi_j)_{j\leq N}\subset H^3_{(0)}$ with $N\in\N^*$. Let $(\rho_{j,k})_{{j,k\leq N}}$ be the 
positive hermitian matrix such that
\begin{align*}
 \pi_N(\Psi)\rho^2\pi_N(\Psi)=\sum_{j,k\leq N}\rho_{j,k}|\psi_j\ra\la\psi_k|.
\end{align*}
There exist $T>0$, $u\in L^2((0,T),\R)$ and $(\theta_{j,k})_{j,k\leq N}$ such that
	\begin{equation*}\begin{split}
	\pi_{N}(\Psi)\ \G_{T}^{u}\rho^1(\G_{T}^{u})^*\ \pi_{N}(\Psi)= \sum_{j,k\leq 
N}e^{i\theta_{j,k}}\rho_{j,k}|\psi_j\ra\la\psi_k|.\\
	\end{split}
	\end{equation*}

\end{theorem}

\begin{proof}
	{\bf 1)} Let $(\psi_j^1)_{j\in\N^*}\subset H^3_{(0)}$ be an orthonormal system made by eigenfunctions of 
$\rho^1$. We have 
\[
\rho^1=\sum_{j=1}^{\infty}l_j|\psi_j^1\ra\la\psi_j^1|,\ \ \ \ \  \ 
\rho^2=\sum_{j=1}^{\infty}l_j|\psi_j\ra\la\psi_j|.
\]
The sequence $(l_j)_{j\in\N^*}\subset \R^+$ corresponds to 
the spectrum of $\rho^1$ and $\rho^2$.
	Now, thanks to Corollary \ref{main_corollary}, there exist $T>0$, $u\in L^2((0,T),\R)$ and $(\theta_j)_{j\leq 
N}$ such that $\pi_N(\Psi)\ \G^{u}_T\psi_j^1=e^{i\theta_j}\pi_N(\Psi)\ \psi_j$ for every $j\leq N$, while 
$\pi_N(\Psi)\ \G^{u}_T\psi_j^1=\pi_N(\Psi)\ \psi_j$ for every $j> N$. Thus, 
	\begin{equation*}\begin{split}
	\pi_N(\Psi)\ \G_T^u\rho^1(\G_T^u)^*\pi_N(\Psi)&=\sum_{j=1}^{N}l_j|e^{i\theta_j}\pi_N(\Psi)\ 
\G_T^u\psi_j^1\ra\la\psi_j^1\G_T^u \pi_N(\Psi)\ e^{i\theta_j}|\\
	&+\sum_{j=N+1}^{\infty}l_j|\pi_N(\Psi)\ \G_T^u\psi_j^1\ra\la\psi_j^1\G_T^u \pi_N(\Psi)\ 
|\\
&=\sum_{j=1}^{\infty}l_j\pi_N(\Psi)\ |\psi_j\ra\la\psi_j|\pi_N(\Psi)=\pi_N(\Psi)\ \rho^2\pi_N(\Psi).
	\end{split}\end{equation*}
{\bf 2)} The second point of the theorem follows from the same arguments of the first one. In particular, the 
statement follows by decomposing $\rho^2$ with respect to $(\psi_{j})_{j\in\N^*}$ as done in 
\eqref{deco_density_orth_basis}. Such 
step provides a positive hermitian matrix $(\rho_{j,k})_{j,k\in\N^*}$. Now, we define $(\psi_{j}^1)_{j\in\N^*}$ 
as the orthonormal system such that \eqref{deco_density_orth_basis1} is valid for the density matrix $\rho^1$. The 
claim is proved by 
simultaneously steering $(\psi_{j}^1)_{j\in\N^*}$ in $(\psi_{j})_{j\in\N^*}$ with respect to the projector 
$\pi(\Psi)$ by using Corollary \ref{main_corollary}.\qedhere
\end{proof}

\section{Conclusion}\label{conclu}
In this manuscript, we study the controllability of the infinite bilinear Schr\"odinger equations 
\eqref{main_equation} at 
the same time $T$, with one unique control $u$ and by projecting onto suitable finite dimensional subspaces of 
$\Hi$. 
The first result of the work is the simultaneous local 
exact controllability of infinite bilinear Schrodinger equations in projection in any positive $T>0$. The 
property is stated by Theorem \ref{theorem_main_local_control} and Proposition \ref{proposition_local_control}. 
Our second achievement is Theorem 
\ref{equivalenza} which shows that the simultaneous global exact 
controllability of the \eqref{main_equation} in projection onto a suitable $N$ dimensional space is equivalent to 
the 
controllability of $N$ problems \eqref{main_equation} (without projecting). Finally, we prove Theorem 
\ref{main_theorem} which states the simultaneous global exact controllability in projection  
of infinite \eqref{main_equation}. The result is guaranteed when the orthogonal 
projector 
is defined by an orthonormal systems verifying a $H^3_{(0)}-$compatibility condition exposed in 
\eqref{compatibilitycondition1}. In conclusion, we rephrase the main results in terms of density matrices.

Here, one could wonder if the techniques developed in this manuscript can be applied to study the controllability 
of infinite \eqref{main_equation} (without projecting). Nevertheless, a direct application is not possible. 
Indeed, 
one of the 
crucial points of our strategy is the possibility of decoupling with a uniform gap the eigenvalues resonances 
appearing in the proof of Theorem \ref{theorem_main_local_control} (see 
Section \ref{introductive_discussion} for 
further details). We obtain such property via perturbation theory techniques thanks to the fact that eigenvalues 
resonances are finite when we project onto finite 
dimensional spaces.

In any case, a possible approach that might lead to the controllability of infinite \eqref{main_equation} is the 
following. 
As already done in our work, one could perturb in order to decouple the eigenvalues resonances 
appearing in the proof of the simultaneous local exact controllability. In such framework, we do not expect to 
have 
a uniform spectral gap and then the Haraux's Theorem \ref{haraux} can not be applied. As a 
consequence, the solvability of the moment problem (such as \eqref{moment_problem_second}) appearing in this proof 
can not be 
achieved in $\ell^2$. Nevertheless, we do not exclude the possibility of proving its solvability in some spaces 
$h^s$ with $s\in [0,1)$ (defined in \ref{spaces}) by using more refined techniques as the Beurling's Theorem 
\cite[Theorem\ \ 9.2]{Ing} (see also \cite[Chapter\ I.2]{expo}). If such result would be valid, then the 
well-posedness of the \eqref{main_equation} can be provided in $H^{3+s}_{(0)}$ by imposing slightly more 
regularity on the 
operator $B$ and we might conclude the proof as done in the current work.

\appendix\section{Moment problem}\label{appendix_moment_problem}

We denote by $\la \cdot,\cdot\ra_{L^2(0,T)}$ the scalar product in $L^2((0,T),\C)$ 
with $T>0.$
\begin{definition}
	Let $(f_k)_{k\in\Z}$ be a family of functions in $L^2((0,T),\C)$ with $T>0$. The family $(f_k)_{k\in\Z}$ 
is 
said to be minimal if and only if $f_k\not\in\overline{\spn\{f_j:j\neq k\}}^{\, L^2}$ for every $k\in\Z.$ 
\end{definition}
	
	\begin{definition} A biorthogonal family to $(f_k)_{k\in\Z}\subset L^2((0,T),\C)$ is a sequence of functions 
$(g_k)_{k\in\Z}$ in $L^2((0,T),\C)$ such that $\la f_k, g_j\ra_{L^2(0,T)}=\delta_{k,j}$ for every $k,j\in\Z.$
	\end{definition}
	
    \begin{remark}\label{unique_bio}
    When $(f_k)_{k\in\Z}$ is minimal, there exists an unique biorthogonal family $(g_k)_{k\in\Z}$ to 
$(f_k)_{k\in\Z}$ belonging to $X:=\overline{\spn\{f_j:j\in\Z\}}^{\, L^2}$. Its existence follows from the fact 
that $(g_k)_{k\in\Z}$ can be 
constructed by setting 
\[
g_k=(f_k-\tilde\pi_kf_k)\|f_k-\pi_kf_k\|^{-2}_{L^2(0,T)},\ \ \ \ \ \ \ \forall k\in\Z
\] 
where 
$\tilde\pi_k$ is the orthogonal projector onto $\overline{\spn\{f_j:j\neq k\}}^{\, L^2}$. The unicity follows as, 
for any biorthogonal family $(g^1_k)_{k\in\Z}$ in $X$, we have $\la g_k 
-g_k^1,f_j\ra_{L^2(0,T)}=0$ for every $j,k\in\Z$, which implies $g_k 
=g_k^1$ for every $k\in\Z.$
       \end{remark}
    
    \begin{remark}\label{minimality}
If a sequence of functions $(f_k)_{k\in\Z}\subset L^2((0,T),\C)$ admits a biorthogonal family 
$(g_k)_{k\in\Z}$, then it is minimal. Indeed, if we assume that there exists $k\in\Z$ such that 
$f_k\in\overline{\spn\{f_j:j\neq k\}}^{\, L^2}$, then the relations $\la 
f_k,g_j\ra_{L^2(0,T)}=0$ for every $j\in\Z\setminus\{k\}$ would imply $\la 
f_k,g_k\ra_{L^2(0,T)}=0$ which is absurd.
    \end{remark}

\begin{definition}
Let $(f_k)_{k\in\Z}$ be a family of functions in $L^2((0,T),\C)$ with $T>0$. The family $(f_k)_{k\in\Z}$ is a 
Riesz basis of $\overline{\spn\{f_j:j\in\Z\}}^{\, L^2}$ if and only if it is isomorphic to an orthonormal system.
\end{definition}

\begin{remark}\label{unique_bio_riesz}    
    Let $(f_k)_{k\in\Z}$ be a Riesz basis of $X:=\overline{\spn\{f_j : j\in\Z\}}^{\, L^2}$. The sequence 
$(f_k)_{k\in\Z}$ is minimal and its biorthogonal family is uniquely defined in 
$X$ 
thanks to Remark \ref{unique_bio}. Finally, this biorthogonal family forms a Riesz basis of $X$.
\end{remark}

Now, we provide an important property on the Riesz basis proved in \cite[Appendix\ B.1]{laurent}.

\begin{proposition}{\cite[Appendix\ B; Proposition\ 19]{laurent}}\label{inequality_prop}
Let $(f_k)_{k\in\Z}$ be a family of functions in $L^2((0,T),\C)$ with $T>0$. The sequence $(f_k)_{k\in\Z}$ 
is a 
Riesz basis of $\overline{\spn\{f_k : k\in\Z\}}^{\, L^2}$ if and only if there exist $C_1,C_1>0$ such that
\begin{equation*}C_1\|{\bf x}\|^2_{\ell^2}\leq\int_0^{ T}\Big|\sum_{k\in\Z}x_kf_k\Big|^2ds\leq 
	C_2\|{\bf x}\|^2_{\ell^2},\ \ \  \ \ \  \ \forall {\bf x}:=(x_k)_{k\in \Z}\in\ell^2(\Z,\C).\end{equation*}
 \end{proposition}

 We are finally ready to present the so-called Haraux's Theorem.

 \begin{theorem}{\cite[Theorem\  4.6]{Ing}}\label{haraux}
Let $(\omega_k)_{k\in \Z}$ be a family of real numbers satisfying the uniform gap condition $\Gi:=\inf_{k\neq 
j}|\omega_{k}-\omega_{j}|>0$. Let $$\Gi':=\sup_{K\subset\Z}\inf_{\underset{k\neq 
j}{k,j\in\Z\setminus K}}|\omega_{k}-\omega_{j}|>0,$$ where $K$ runs over the finite subsets of $\Z$. For every 
bounded interval 
$|I|>\frac{2\pi}{\Gi'}$, there exist $C_1,C_2>0$ such that
\[
C_1\sum_{k\in \Z}|x_k|^2\leq\int_{I}|u(t)|^2dt\leq C_2\sum_{k\in 
\Z}|x_k|^2,
\]
for every $u(t)=\sum_{k\in \Z}x_k e^{i\omega_kt}$
with $(x_k)_{k\in\Z}\in \ell^2(\Z,\C)$.
\end{theorem}

The following corollary follows from the Haraux's Theorem and provides the solvability of suitable moment 
problems as \eqref{first_moment_problem} and \eqref{moment_problem_second}.

 \begin{corollary}\label{solvability_moment_problem}
Let $(\lambda_k)_{k\in\N^*}$ be an ordered sequence of real numbers such that $\lambda_1=0$ and $\Gi:=\inf_{k\neq 
j}|\lambda_{k}-\lambda_{j}|>0$. Let 
$$\Gi':=\sup_{K\subset\N^*}\inf_{\underset{k\neq j}{k,j\in\N^*\setminus K}}|\lambda_{k}-\lambda_{j}|,$$ where $K$ 
runs over the finite subsets of $\N^*$.  
Fixed $T>2\pi/\Gi'$, for every $(x_k)_{k\in\N^*}\in \ell^2(\C)$, there exists $u\in L^2((0,T),\R)$ such that
\begin{equation}\begin{split}\label{mome_appen} x_k=\int_0^Tu(s)e^{-i\lambda_k s}ds,\ \ \ \ \ \ \ \ \forall k\in 
\N^*.\\
\end{split}\end{equation}
\end{corollary}
\begin{proof}
For $k\in\N^*$, we call $\omega_k=\lambda_{k}$, while we impose $\omega_k=-\lambda_{-k}$ for 
$-k\in\N^*\setminus\{1\}$. 	We call $\Z^*=\Z\setminus\{0\}$. 
	The sequence $(\omega_k)_{k\in \Z^*\setminus\{-1\}}$ satisfies the hypotheses of Theorem \ref{haraux}
for 
\[
\sup_{K\subset\Z^*\setminus\{-1\}}\inf_{\underset{k\neq 
j}{k,j\in(\Z^*\setminus\{-1\})\setminus K}}|\omega_{k}-\omega_{j}|=\Gi',
\]
 where $K$ runs over the finite subsets 
of 
$\Z^*\setminus\{-1\}$. Proposition \ref{inequality_prop} and Theorem \ref{haraux} ensure that the 
sequence $(e^{i\omega_kt})_{k\in\Z^*\setminus\{-1\}}$ is a Riesz basis of  
\[
X:=\overline{\spn\{e^{i\omega_kt} : 
k\in\Z^*\setminus\{-1\}\}}^{\, L^2}.
\] 
Thanks to Remark \ref{unique_bio_riesz}, its unique biorthogonal 
family $(v_k)_{k\in\Z^*\setminus\{-1\}}$ in $X$ is also a Riesz basis of $X$. Thanks to Proposition 
\ref{inequality_prop}, there exist $C_1,C_2>0$ such that 
\[
C_1\sum_{k\in 
\Z^*\setminus\{-1\}}|x_k|^2\leq\|u\|_{L^2(0,T)}^2\leq C_2\sum_{k\in \Z^*\setminus\{-1\}}|x_k|^2,
\] 
with $u(t)=\sum_{k\in  \Z^*\setminus\{-1\}}x_k v_k(t)$ and 
$(x_k)_{k\in\Z^*\setminus\{-1\}}\in\ell^2(\Z^*\setminus\{-1\},\C).$
Now, 
$$u=\sum_{k\in\Z^*\setminus\{-1\}}v_k \la e^{i\omega_kt},u\ra_{L^2(0,T)}$$
since $(e^{i\omega_kt})_{k\in\Z^*\setminus\{-1\}}$ and $(v_k)_{k\in\Z^*\setminus\{-1\}}$ are reciprocally 
biorthogonal. Hence,
\[
C_1\sum_{k\in \Z^*\setminus\{-1\}}|\la 
e^{i\omega_kt},u\ra_{L^2(0,T)}|^2\leq\|u\|_{L^2(0,T)}^2\leq C_2\sum_{k\in \Z^*\setminus\{-1\}}|\la 
e^{i\omega_kt},u\ra_{L^2(0,T)}|^2.
\]
The last relation yields the invertibility of the map 
\[
F:u\in X\longmapsto \left(\la 
e^{i\omega_kt},u\ra_{L^2(0,T)}\right)_{k\in\Z^*\setminus\{-1\}}\in\ell^2(\Z^*\setminus\{-1\},\C).
\]
Fixed $(x_k)_{k\in\N^*}\in \ell^2(\C)$. We call $(\widetilde x_k)_{k\in \Z^*\setminus\{-1\}}\in 
\ell^2(\Z^*\setminus\{-1\},\C)$ the sequence such that $\widetilde x_k=x_{k}$ for $k\in\N^*$, while 
$\widetilde 
x_k=\overline 
x_{-k}$ for $-k\in \N^*\setminus\{1\}$. 
	For $T>2\pi/\Gi'$, the invertibility of the map $F$ ensures the existence of $u\in L^2((0,T),\C)$ 
such that
\[
\widetilde 
x_k=\int_0^Tu(s)e^{-i\omega_k s}ds
\]
for every $k\in \Z^*\setminus\{-1\}$. Thus, \begin{equation*}\begin{split}\begin{cases}
	x_k=\int_0^Tu(s)e^{-i\lambda_k s}ds=\int_0^T\overline{u}(s)e^{-i\lambda_k s}ds,\ \ \ \ & \forall  
k\in\N^*\setminus\{1\},\\
	x_1=\int_0^Tu(s)ds.\\
	\end{cases}
	\end{split}\end{equation*}
	Finally, if $x_{1}\in\R$, then \eqref{mome_appen} is valid with respect to a function $u$ which is real. 
\qedhere
\end{proof}

\section{Analytic Perturbation}\label{appendix_analitics}

Let us consider the problem \eqref{main_equation_modified} and the eigenvalues $(\lambda_j^{u_0})_{j\in\N^*}$ of 
the operator 
$A+u_0B$. When $B$ is a bounded symmetric operator satisfying Assumptions I and $A=-\Delta$ is the Laplacian with 
Dirichlet type boundary conditions $D(A)=H^2((0,1),\C)\cap H^1_0((0,1),\C),$
	thanks to \cite[Theorem\ VII.2.6]{kato} and \cite[Theorem\ VII.3.9]{kato}, the following proposition 
follows.
	
	\begin{proposition}\label{analiticity_perturbation}
		Let $B$ satisfy Assumptions I. There exists a neighborhood $D(0)$ of $u=0$ in $\R$ small enough where the 
maps $u\mapsto\lambda_j^{u}$ are analytic for every $ j\in\N^*$.
	\end{proposition}
	
	The next lemma proves the existence of perturbations, which do not shrink too much the eigenvalues gaps. 

	\begin{lemma}\label{bound_perturbed_resolvent}
		Let $B$ satisfy Assumptions I. There exists a neighborhood $D(0)$ in $\R$ of $u=0$ such that, for each 
$u_0\in D(0)$, there exists $r>0$ such that, for every $ j\in\N^*$,
\[
\mu_j:=\frac{\lambda_j+\lambda_{j+1}}{2}\in\rho(A+u_0B),\ \ \ \ \ \ \iii(A+u_0B-\mu_j)^{-1}\iii\leq r.
\]
	\end{lemma}
	\begin{proof}
		Let $D(0)$ be the neighborhood provided by Proposition \ref{analiticity_perturbation}. We know $(A-\mu_j)$ 
is invertible in 
a bounded operator and $\mu_j\in\rho(A)$ (resolvent set of $A$). Let 
$\delta:=\min_{j\in\N^*}|\lambda_{j+1}-\lambda_j|.$ We know that $\iii(A-\mu_j)^{-1}\iii 
\leq\sup_{k\in\N^*}\frac{1}{|\mu_j-\lambda_k|}=\frac{2}{|\lambda_{j+1}-\lambda_j|}\leq\frac{2}{\delta}.$ Thus, for 
$u_0\in D(0)$, 
\[
\iii(A-\mu_j)^{-1}u_0B\iii\leq|u_0|\iii(A-\mu_j)^{-1}\iii\iii B\iii\leq \frac{2}{\delta}|u_0|\iii 
B\iii
\]
and if $|u_0|\leq\frac{\delta(1-\epsilon)}{2\iii B\iii}$ for $\epsilon\in(0,1),$ then 
$\iii(A-\mu_j)^{-1}u_0B\iii\leq 1-\epsilon.$ The operator $(A+u_0B-\mu_j)$ is invertible and 
$\iii(A+u_0B-\mu_j)^{-1}\iii\leq\frac{2}{\delta\epsilon}$ as 
$\|(A+u_0B-\mu_j)\psi\|_{L^2}\geq\|(A-\mu_j)\psi\|_{L^2}-\| 
u_0B\psi\|_{L^2}\geq\frac{\delta}{2}\|\psi\|_{L^2}-\frac{\delta(1-\epsilon)}{2}\|\psi\|_{L^2}$ for every $\psi\in 
D(A)$. The parameter $r$ stated in the lemma corresponds to $2/(\delta\epsilon)$, while the neighborhood is 
$\{u_0\in D(0)\ :\ |u_0|\leq{\delta(1-\epsilon)}/{(2\iii B\iii)}\}$.\qedhere
		\end{proof}

	\begin{lemma}\label{invertibility_perturbation}
		Let $B$ satisfy Assumptions I and $P_{\phi_k}^{\bot}$ be the projector onto the orthogonal space of 
$\phi_k$. There exists a neighborhood $D(0)$ of $0$ in $\R$ such that 
\[
(A+u_0P_{\phi_k}^{\bot}B-\lambda_k^{u_0})
\]
is invertible with bounded inverse from $D(A)\cap \phi_k^{\bot}$ to $\phi_k^{\bot}$ for every $ u_0\in D(0)$ and $ 
k\in\N^*$. 
	\end{lemma}
	\begin{proof}
		Let $D(0)$ be the neighborhood provided by Lemma \ref{bound_perturbed_resolvent}. For any $u_0\in D(0)$, 
one can consider the decomposition 
$(A+u_0P_{\phi_k}^{\bot}B-\lambda_k^{u_0})=(A-\lambda_k^{u_0})+u_0P_{\phi_k}^{\bot}B.$ The operator 
$A-\lambda_k^{u_0}$ is invertible with bounded inverse when it acts on the orthogonal space of $\phi_k$ and we 
estimate $\iii ((A-\lambda_k^{u_0})\big|_{\phi_k^{\bot}})^{-1}u_0P_{\phi_k}^{\bot}B\iii.$ However, for every 
$\psi\in D(A)\cap Ran(P_{\phi_k}^{\bot})$ such that $\|\psi\|_{L^2} =1$, we have 
\[
\|(A-\lambda_k^{u_0})\psi\|_{L^2} 
\geq\min\{|\lambda_{k+1}-\lambda_k^{u_0}|,|\lambda_k^{u_0}-\lambda_{k-1}|\}\|\psi\|_{L^2}.
\]
		Let $\delta_k:=\min\big\{|\lambda_{k+1}-\lambda_k^{u_0}|,|\lambda_k^{u_0}-\lambda_{k-1}|\big\}.$
		Thanks to Lemma \ref{bound_perturbed_resolvent}, for $|u_0|$ small enough, 
$\lambda_k^{u_0}\in\left(\frac{\lambda_{k-1}+\lambda_{k}}{2},\frac{\lambda_{k}+\lambda_{k+1}}{2}\right)$
		and then
		
\begin{equation*}\begin{split}\delta_k&\geq\min\Big\{\Big|\lambda_{k+1}-\frac{\lambda_{k}+\lambda_{k+1}}{2}\Big|,
\Big|\frac{\lambda_{k-1}+\lambda_{k}}{2}-\lambda_{k-1}\Big|\Big\} 
\geq\frac{(2k-1)\pi^2}{2}>k.\end{split}\end{equation*}
		Afterwards, $$\iii 
((A-\lambda_k^{u_0})\big|_{\phi_k^{\bot}})^{-1}u_0P_{\phi_k}^{\bot}B\iii\leq\frac{1}{\delta_k}|u_0|\iii B\iii$$ 
and, if $|u_0|\leq(1-r)\frac{\delta_k}{\iii B\iii}\leq\frac{(1-r)}{\iii B\iii}$ for $r\in(0,1)$, then it follows 
$$\iii ((A-\lambda_k^{u_0})\big|_{\phi_k^{\bot}})^{-1}u_0P_{\phi_k}^{\bot}B\iii\leq(1-r)<1.$$ The operator 
$A_k:=(A-\lambda_k^{u_0}+u_0P_{\phi_k}^{\bot}B)$ is invertible when it acts on the orthogonal space of $\phi_k$ 
and, for every $\psi\in D(A)$ and $r=\frac{1}{2}$,
		\begin{equation*}\begin{split}
		\| A_k\psi\|_{L^2}&\geq\| (A-\lambda_k^{u_0})\psi\|_{L^2}-\| 
u_0P_{\phi_k}^{\bot}B\psi\|_{L^2}\\
&\geq\delta_k\|\psi\|_{L^2}-\iii 
u_0P_{\phi_k}^{\bot}B\iii\|\psi\|_{L^2}\geq\frac{1}{2}\|\psi\|_{L^2}.\\
		\end{split}\end{equation*} 
		In conclusion, $\iii ((A-\lambda_k^{u_0}+u_0P_{\phi_k}^{\bot}B)\big|_{\phi_k^{\bot}}
		)^{-1}\iii\leq 2$ for every $k\in\N^*.$ \qedhere 
		\end{proof}

	\begin{lemma}\label{perturb_eigenvalues}
		Let $B$ satisfy Assumptions I. There exists a neighborhood $D(0)$ of $0$ in $\R$ such that, for any 
$u_0\in 
D(0)$, we have $\lambda_j^{u_0}\neq 0$ and there exist two constants $C_1,C_2>0$ such that
\[
C_1\lambda_j\leq\lambda_j^{u_0}\leq C_2\lambda_j,\ \ \ \ \  \ \ \forall j\in\N^*.
\]
	\end{lemma}
	\begin{proof}
		Let $u_0\in D(0)$ for $D(0)$ the neighborhood provided by Lemma \ref{invertibility_perturbation}.
		We decompose the eigenfunction $\phi_j^{u_0}=a_j\phi_j+\eta_j,$ where $a_j$ is an orthonormalizing 
constant and $\eta_j$ is orthogonal to $\phi_j$. Hence $\lambda_k^{u_0}\phi_k^{u_0}=(A+u_0B)(a_k\phi_k+\eta_k)$ 
and $
		\lambda_k^{u_0}a_k\phi_k+\lambda_k^{u_0}\eta_k=Aa_k\phi_k+A\eta_k+u_0Ba_k\phi_k+u_0B\eta_k$. By projecting 
onto the orthogonal space of $\phi_k$,
		\[
		\lambda_k^{u_0}\eta_k=A\eta_k+u_0P_{\phi_k}^{\bot}Ba_k\phi_k+u_0P_{\phi_k}^{\bot}B\eta_k.
		\]
		However, Lemma 
\ref{invertibility_perturbation} ensures that $A+u_0P_{\phi_k}^{\bot}B-\lambda_k^{u_0}$ is invertible with bounded 
inverse when it acts on the orthogonal space of $\phi_k$ and then
		
\begin{equation}\label{rest_perturbation}\eta_k=-a_k((A+u_0P_{\phi_k}^{\bot}B-\lambda_k^{u_0})\big|_{\phi_k^{\bot}}
)^{-1}u_0P_{ \phi_k}^{\bot}B\phi_k.\end{equation}
Now,		\begin{equation*}\begin{split}
		\lambda_j^{u_0}&=\la 
a_j\phi_j+\eta_j,(A+u_0B)(a_j\phi_j+\eta_j)\ra_{L^2}=|a_j|^2\lambda_j+u_0\la a_j\phi_j,Ba_j\phi_j\ra_{L^2}\\
		&+\la 
a_j\phi_j,(A+u_0B)\eta_j\ra_{L^2}+\la\eta_j,(A+u_0B)a_j\phi_j\ra_{L^2}+\la\eta_j,(A+u_0B)\eta_j\ra_{L^2}.\\
		\end{split}\end{equation*}
		By using the relation \eqref{rest_perturbation},
		\begin{equation*}\begin{split}
\la\eta_j,(A+u_0B)\eta_j\ra_{L^2}&=\la\eta_j,(A+u_0P_{\phi_k}^{\bot}B-\lambda_j^{u_0})\eta_j\ra_{L^2}+\lambda_j^{ 
u_0}\|\eta_j\|_{L^2}^2\\
&=\lambda_j^{u_0}\|\eta_j\|_{L^2}^2-a_j \big\la\eta_j,u_0P_{\phi_j}^{\bot}B\phi_j\big\ra_{L^2}.\\ 
		\end{split}\end{equation*}
		However, $\la\phi_j,(A+u_0B)\eta_j\ra_{L^2}=u_0\la\phi_j,B\eta_j\ra_{L^2}=u_0\la 
P_{\phi_j}^{\bot}B\phi_j,\eta_j\ra_{L^2}$	and $\la\eta_j,(A+u_0B)\phi_j\ra_{L^2}=u_0\la 
\eta_j,P_{\phi_j}^{\bot}B\phi_j\ra_{L^2}.$ Thus, the last relations yields
		\begin{equation}\begin{split}\label{decomposition_perturbed_eigenvalue}
		\lambda_j^{u_0}&=|a_j|^2\lambda_j+u_0|a_j|^2B_{j,j}+\lambda_j^{u_0}\|\eta_j\|_{L^2}^2+u_0\overline{a_j}\la 
P_{\phi_j}^{\bot}B\phi_j,\eta_j\ra_{L^2}.\\ 
		\end{split}\end{equation}
		One can notice that $|a_j|\in[0,1]$ and $\|\eta_j\|_{L^2}$ are uniformly bounded in $j$. We show that the 
first accumulates at $1$ and the second at $0$. Indeed, from the proof of Lemma 
\eqref{invertibility_perturbation} and the relation \eqref{rest_perturbation}, there exists $C_1>0$ such that
		\begin{equation}\label{upper_bound_resto}
		\begin{split}
\|\eta_j\|_{L^2}^2&\leq|u_0|^2\iii((A+u_0P_{\phi_j}^{\bot}B-\lambda_j^{u_0})\big|_{\phi_j^{\bot}})^{-1}
\iii^2|a_j|^2\|B\phi_j\|_{L^2}^2	\leq\frac{C_1}{j^2}\\
		\end{split}\end{equation}
		for $r\in (0,1)$, which implies that $\lim_{j\rightarrow\infty}\|\eta_j\|_{L^2}=0.$ Afterwards, by 
contradiction, if $|a_j|$ does not converge to $1$, then there exists $(a_{j_k})_{k\in\N^*}$ a subsequence of 
$(a_{j})_{j\in\N^*}$ such that $|a_{j_\infty}|:=\lim_{k\rightarrow\infty}|a_{j_k}| \in [0,1)$. Now, we have
\[
1=\lim_{k\rightarrow\infty}\|\phi_{j_k}^{u_0}\|_{L^2} 
\leq\lim_{k\rightarrow\infty}|a_{j_k}|\|\phi_{j_k}\|_{L^2}+\|\eta_{j_k}\|_{L^2}=\lim_{k\rightarrow\infty}|a_{j_k}
|+\|\eta_{j_k}\|_{L^2}=|a_{j_\infty}|<1
\]
that is absurd. Then, $\lim_{j\rightarrow\infty}|a_j|=1$. From \eqref{decomposition_perturbed_eigenvalue}, it 
follows that there exist two constants $C_1,C_2>0$ such that, for each $j\in\N^*$, 
$C_1\lambda_j\leq\lambda_j^{u_0}\leq C_2\lambda_j$ for $|u_0|$ small enough.
		The relation also implies that $\lambda_j^{u_0}\neq 0$ for every $j\in\N^*$ and $|u_0|$ small 
enough.\qedhere
			\end{proof}

	\begin{lemma}\label{perturbation_mixing_eigenspaces}
		Let $B$ satisfy Assumptions I. For every $N\in\N^*$, there exist a neighborhood $D(0)$ of $0$ in $\R$ and 
$\widetilde C_N >0$ such that, for any $u_0\in D(0)$, we have 
\[
|\la\phi_k^{u_0},B 
\phi_j^{u_0}\ra_{L^2}|\geq\frac{\widetilde C_N }{k^3},\ \ \ \ \ \ \forall k,j\in\N^*,\ j\leq N.
\]
\end{lemma}
\begin{proof}
We start by choosing $k\in\N^*$ such that $k\neq j$ and $u_0\in D(0)$ for $D(0)$ the neighborhood provided by 
Lemma \ref{perturb_eigenvalues}. Thanks to Assumptions I, we have
	\begin{equation}\label{primo}\begin{split}
	&|\la\phi_k^{u_0},B\phi_j^{u_0}\ra_{L^2}|=|\la a_k\phi_k+\eta_k,B(a_j\phi_j+\eta_j)\ra_{L^2}|\\
	&\geq 
C_N\frac{\overline{a_k}a_j}{k^3}-\big|\overline{a_k}\la\phi_k,B\eta_j\ra_{L^2}+a_j\la\eta_k,B\phi_j\ra_{L^2}
+\la\eta_k,B\eta_j\ra_{L^2}\big|.\\
	\end{split}
	\end{equation}

	\needspace{3\baselineskip}
	
	\noindent
	{\bf 1) Expansion of $\la\eta_k,B\phi_j\ra_{L^2}$, $\la\phi_k,B\eta_j\ra_{L^2}$ and 
$\la\eta_k,B\eta_j\ra_{L^2}$:  }	Thanks to \eqref{rest_perturbation}, 
$$\la\eta_k,B\phi_j\ra_{L^2}=\la-a_k((A+u_0P_{\phi_k}^{\bot}B-\lambda_k^{u_0})\big|_{\phi_k^{\bot}})^{-1}u_0P_{
\phi_k}^{\bot}B\phi_k,P_{\phi_k}^{\bot}B\phi_j\ra_{L^2}$$ for every $k\in\N^*$ and $j\leq N$, while the operator 
$\big((A+u_0P_{\phi_k}^{\bot}B-\lambda_k^{u_0})\big|_{\phi_k^{\bot}}\big)^{-1}$ corresponds to
	$$((A-\lambda_k^{u_0})P_{\phi_k}^{\bot})^{-1}\sum_{n=0}^{\infty}\big(u_0 
((A-\lambda_k^{u_0})P_{\phi_k}^{\bot})^{-1} P_{\phi_k}^{\bot}BP_{\phi_k}^{\bot}\big)^n$$
	for $|u_0|$ small enough. For $$M_k:=\sum_{n=0}^{\infty}\big(u_0 ((A-\lambda_k^{u_0})P_{\phi_k}^{\bot})^{-1} 
P_{\phi_k}^{\bot}B\big)^nP_{\phi_k}^{\bot},$$ we have
\[
\la\eta_k,B\phi_j\ra_{L^2}=-u_0\la a_kM_kB\phi_k, 
((A-\lambda_k^{u_0})P_{\phi_k}^{\bot})^{-1}P_{\phi_k}^{\bot}B\phi_j\ra_{L^2}.
\]
Thanks to $B:D(A)\rightarrow D(A)$, for every $k\in\N^*$ and $j\leq N$,
	\begin{equation*}
	\begin{split}
	((A-\lambda_k^{u_0})P_{\phi_k}^{\bot})^{-1}P_{\phi_k}^{\bot}B\phi_j
&=P_{\phi_k}^{\bot}B((A-\lambda_k^{u_0})P_{\phi_k}^{\bot})^{-1}\phi_j\\
&-\big[P_{\phi_k}^{\bot}B,((A-\lambda_k^{u_0}
)P_
{\phi_k}^{\bot})^{-1}P_{\phi_k}^{\bot}\big]\phi_j\\
&=P_{\phi_k}^{\bot}B((A-\lambda_k^{u_0})P_{\phi_k}^{\bot})^{-1}\phi_j\\
&-((A-\lambda_k^{u_0})P_{\phi_k}^{\bot})^{-1}
P_
{\phi_k}^{\bot}[B,A]((A-\lambda_k^{u_0})P_{\phi_k}^{\bot})^{-1}\phi_j.\\
	\end{split}
	\end{equation*}
	For $\widetilde B_k:=((A-\lambda_k^{u_0})P_{\phi_k}^{\bot})^{-1}P_{\phi_k}^{\bot}[B,A],$ we have 
$$((A-\lambda_k^{u_0})P_{\phi_k}^{\bot})^{-1}P_{\phi_k}^{\bot}B\phi_j=P_{\phi_k}^{\bot}(B+\widetilde 
B_k)(\lambda_j-\lambda_k^{u_0})^{-1}\phi_j$$ and, for every $k\in\N^*$ and $j\leq N$,
	\begin{equation}\label{secondo}\begin{split}
	\la\eta_k,B\phi_j\ra_{L^2}&=-\frac{u_0}{\lambda_j-\lambda_k^{u_0}}\la a_k M_kB\phi_k,(B+\widetilde 
B_k)\phi_j\ra_{L^2}.\\
	\end{split}
	\end{equation}
	
	\noindent
	For every $k\in\N^*$ and $j\leq N$, we obtain
	\begin{equation}
	\label{secondo_2}
	\begin{split}
	&|\la\eta_k,B\eta_j\ra_{L^2}|=|\la B\eta_k,\eta_j\ra_{L^2}|=|\la u_0 a_k 
B((A-\lambda_k^{u_0})P_{\phi_k}^{\bot})^{-1}M_kB\phi_k,\\
	&u_0 a_j ((A-\lambda_j^{u_0})P_{\phi_j}^{\bot})^{-1}M_jB\phi_j\ra_{L^2}\Big|=\Big|\frac{a_j\overline{a_k} 
u_0^2}{\lambda_k-\lambda_j^{u_0}}\big\la \phi_k,L_{k,j}\phi_j\big\ra_{L^2}\Big|\\
	\end{split}\end{equation}
	with 
$L_{k,j}:=(A-\lambda_j^{u_0})BM_k((A-\lambda_k^{u_0})P_{\phi_k}^{\bot})^{-1}P_{\phi_k}^{\bot}B((A-\lambda_j^{u_0}
)P_{\phi_j}^{\bot})^{-1}M_jB.$	Now, there exists $\epsilon>0$ such that $|a_l|\in (\epsilon,1)$ for every 
$l\in\N^*$. Thanks to \eqref{secondo}, \eqref{secondo_2} and \eqref{primo}, there exists $\widehat C_N$ such 
that
	\begin{equation}\label{terzo}\begin{split}
	|&\la\phi_k^{u_0},B\phi_j^{u_0}\ra_{L^2}|\geq \widehat 
C_N\frac{1}{k^3}-\Big|\frac{u_0}{\lambda_j-\lambda_k^{u_0}}\la M_kB\phi_k,(B+\widetilde B_k)\phi_j\ra_{L^2}\Big|\\
	&-\Big|\frac{u_0}{\lambda_k-\lambda_j^{u_0}}\la (B+\widetilde 
B_j)\phi_k,M_jB\phi_j\ra_{L^2}\Big|-\Big|\frac{u_0^2}{\lambda_k-\lambda_j^{u_0}}\big\la 
\phi_k,L_{k,j}\phi_j\big\ra_{L^2}\Big|.\\
	\end{split}
	\end{equation}

	\needspace{3\baselineskip}
	
	\noindent
	{\bf 2) Features of the operators $M_k$, $\widetilde B_k$ and $L_{k,j}$.}	Each $M_k$ for $k\in\N^*$ is 
uniformly bounded in $L(H^2_{(0)},H^2_{(0)})$ when $|u_0|$ is small enough such that $$\iii u_0 
((A-\lambda_k^{u_0})P_{\phi_k}^{\bot})^{-1} P_{\phi_k}^{\bot}BP_{\phi_k}^{\bot}\iii_{L(H^2_{(0)})}<1.$$
	The definition of $\widetilde B_k$ implies that $$\widetilde 
B_kP_{\phi_k}^{\bot}=((A-\lambda_k^{u_0})P_{\phi_k}^{\bot})^{-1}P_{\phi_k}^{\bot}B(A-\lambda_k^{u_0})P_{\phi_k}^{
\bot}-P_{\phi_k}^{\bot}BP_{\phi_k}^{\bot}.$$ Hence, the operators $\widetilde B_k$ are uniformly bounded in $k$ in 
$L\big(H^2_{(0)}\cap Ran(P_{\phi_k}^{\bot}),H^2_{(0)}\cap Ran(P_{\phi_k}^{\bot})\big).$
	Third, one can notice that $$B((A-\lambda_j^{u_0})P_{\phi_j}^{\bot})^{-1}M_jB\in L(H^2_{(0)},H^2_{(0)})$$ for 
every $j\in\N^*.$ Then, for every $k\in\N^*$ and $j\leq N$,
	\begin{equation*}\begin{split}
	(A-\lambda_j^{u_0})BM_k((A-\lambda_k^{u_0})P_{\phi_k}^{\bot})^{-1}P_{\phi_k}^{\bot}&= 
(A-\lambda_j^{u_0})B((A-\lambda_k^{u_0})P_{\phi_k}^{\bot})^{-1}\\
	&\sum_{n=0}^{\infty}\big(u_0 P_{\phi_k}^{\bot}B((A-\lambda_k^{u_0})P_{\phi_k}^{\bot})^{-1} 
\big)^nP_{\phi_k}^{\bot}\\
&= 
(A-\lambda_j^{u_0})((A-\lambda_k^{u_0})P_{\phi_k}^{\bot})^{-1}P_{\phi_k}^{\bot}(\widetilde B_k+B)\widetilde M_k\\
	\end{split}\end{equation*}
	with $$\widetilde M_k:=\sum_{n=0}^{\infty}\big(u_0 
P_{\phi_k}^{\bot}B((A-\lambda_k^{u_0})P_{\phi_k}^{\bot})^{-1} \big)^nP_{\phi_k}^{\bot}.$$ Finally, the operators 
$\widetilde M_k$ are uniformly bounded in $L(H^2_{(0)},H^2_{(0)})$ as $M_k$. Hence $L_{k,j}$ are uniformly bounded 
in $L(H^2_{(0)},H^2_{(0)})$.

	Let $(F_l)_{l\in\N^*}$ be an infinite uniformly bounded family of operators in $L(H^2_{(0)},H^2_{(0)})$. For 
every $l,j\in\N^*$, there exists $c_{l,j}>0$ such that $$\sum_{k=1}^\infty |k^2\la \phi_k, F_l \phi_j 
\ra_{L^2}|^2<\infty,\ \ \ \Longrightarrow\ \ \ \ |\la \phi_k, F_l \phi_j \ra_{L^2}|\leq \frac{c_{l,j}}{k^2}$$ for 
every 
$k\in\N^*.$ Now, the constant $c_{l,j}$ can be assumed uniformly bounded in $l$ since, for every $k,j\in\N^*$,
	\begin{equation*}
	\begin{split}
	\sup_{l\in\N^*}|k^2\la\phi_k, F_l\phi_j\ra_{L^2}|^2&\leq\sup_{l\in\N^*}\sum_{m\in\N^*}|m^2\la\phi_m, 
F_l\phi_j\ra_{L^2}|^2	\leq \sup_{l\in\N^*}\|F_l\phi_j\|_{(2)}^2<\infty.\\
	\end{split}
	\end{equation*}

	\noindent
	Thus, for every infinite uniformly bounded family of operators $(F_l)_{l\in\N^*}$ in $L(H^2_{(0)},H^2_{(0)})$ 
and for every $j\in\N^*$, there exists a constant $c_j$ such that
	\begin{equation}\label{quarto}
	|\la \phi_k, F_l \phi_j \ra_{L^2}|\leq \frac{c_j}{k^2},	\ \ \ \ \ \ \ \forall k,l\in\N^*.\\
	\end{equation}

	\needspace{3\baselineskip}
	\noindent
	{\bf 3) Conclusion.} We know that $|\lambda_j-\lambda_k^{u_0}|^{-1}$ and $|\lambda_k-\lambda_j^{u_0}|^{-1}$ 
asymptotically behave as $k^{-2}$ thanks to Lemma \ref{perturb_eigenvalues}. 
	From the previous point, the families of operators $(B M_k(B+\widetilde B_k))_{k\in\N^*}$, 
$(L_{k,j})_{k\in\N^*}$ are uniformly bounded in $L(H^2_{(0)},H^2_{(0)})$ and $B M_j(B+\widetilde B_j)\in 
L(H^2_{(0)},H^2_{(0)})$ for every $1\leq j\leq N$. Hence, we use the relation \eqref{quarto} in \eqref{terzo} 
and there exist $C_1,C_2,C_3,C_4>0$ depending on $j\in\N^*$ such that, for $|u_0|$ small enough and $k\in\N^*$ 
large enough,
	\begin{equation}\label{quinto}\begin{split}
	|\la\phi_k^{u_0},B\phi_j^{u_0}\ra_{L^2}|	\geq&\ \widehat 
C_N\frac{1}{k^3}-\frac{C_1|u_0|}{|\lambda_j-\lambda_k^{u_0}|k^2}-\frac{C_2|u_0|}{|\lambda_k-\lambda_j^{u_0}|k^2}\\
&
-\frac{C_3|u_0|^2}{|\lambda_k-\lambda_j^{u_0}|k^2}\geq C_4\frac{1}{k^3}.\\
	\end{split}
	\end{equation}
	Let $K\in\N^*$ be such that $|\la\phi_k^{u_0}(T),B\phi_j^{u_0}(T)\ra_{L^2}|\geq C_4\frac{1}{k^3}$ for every 
$k> 
K.$ For $j\in\N^*$, the zeros of the analytic map $u_0\mapsto 
(|\la\phi_k^{u_0}(T),B\phi_j^{u_0}(T)\ra_{L^2}|)_{k\leq K}\in \R^{K}$ are discrete. Then, for $|u_0|$ small 
enough, $|\la\phi_k^{u_0}(T),B\phi_j^{u_0}(T)\ra_{L^2}|\neq 0$ for every $k\leq K.$ Thus, for every $j\in\N^*$ and 
$|u_0|$ small enough, there exists $C_j>0$ such that $|\la\phi_k^{u_0}(T),B\phi_j^{u_0}(T)\ra_{L^2}|\geq 
\frac{C_j}{k^3}$ for every $k\in\N^*.$ In conclusion, the claim is achieved for every $k\in\N^*$ and $j\leq N$ 
with 
$\widetilde C_N=\min\{C_j:\ j\leq N\}$.
	\qedhere
\end{proof}

	\begin{lemma}\label{perturnated_norm_equivalence}
		Let $B$ satisfy Assumptions I. There exists a neighborhood $D(0)$ of $0$ in $\R$ such that, for any 
$u_0\in 
D(0)$, there exist  $C_1,C_2>0$ such that
		
\[
C_1\Big(\sum_{j=1}^{\infty}\big||\lambda_j^{u_0}|^\frac{3}{2}\la\phi_j^{u_0},\cdot\ra_{L^2}\big|^2\Big)^{\frac{1}
{2}}
		\leq\|\cdot\|_{(3)}\leq 
C_2\Big(\sum_{j=1}^{\infty}\big||\lambda_j^{u_0}|^\frac{3}{2}\la\phi_j^{u_0},\cdot\ra_{L^2}\big|^2\Big)^{\frac{1}{2
}}.
\]
	\end{lemma}
	\begin{proof}
		Let $D(0)$ be the neighborhood provided by Lemma \ref{perturb_eigenvalues}. For $|u_0|$ small enough, we 
prove that there exist $C_1>0$ such that $\||A+u_0B|^\frac{s}{2}\psi\|_{L^2}\leq C_1\| 
|A|^\frac{s}{2}\psi\|_{L^2}$ for $s=3$. We start with $s=4$ and we recall that $B\in {L(H^2_{(0)})}$ thanks to 
Remark \ref{osss_closed}. 
		For any $\psi\in H^4_{(0)}$, there exists $C_2>0$ such that
\begin{equation*}\begin{split}
\|(A+u_0B)^2\psi\|_{L^2}\leq&\ \| A^2\psi\|_{L^2}+|u_0|^2\|B^2\psi\|_{L^2}\\
&+|u_0|\|A\psi\|_{L^2}(\iii 
B\iii_{(2)}+\iii B\iii)\leq C_2\| |A|^2\psi\|_{L^2}.\\
\end{split}\end{equation*}
Classical interpolation arguments (see for instance the proof of \cite[Lemma\ 
1]{nabile}) imply the 
validity of the relation also for $s=3$. There exists $C>0$ such that $$\| \psi\|_{\widetilde 
H^3_{(0)}}=\||A+u_0B|^\frac{3}{2}\psi\|_{L^2}\leq C\||A|^\frac{3}{2}\psi\|_{L^2}=C \| \psi\|_{ H^3_{(0)}}$$ for 
every $\psi\in H^3_{(0)}$. Now, $H^2_{(0)}=D(|A|)=D(|A+u_0B|)=\widetilde H^2_{(0)}$ and  
$B:H^2_{(0)}\longrightarrow H^2_{(0)}$. The arguments of Remark \ref{osss_closed} imply that 
$B\in L(\widetilde H^2_{(0)})$ and the opposite inequality follows as above
		from the decomposition $A=(A+u_0B)-u_0B$.\qedhere
	\end{proof}
	\begin{remark}\label{perturbated_norm_equivalence_remark}
		Let $B$ satisfy Assumptions I. The techniques of the proof of Lemma \ref{perturnated_norm_equivalence} 
also allow to prove that, 
for $s\in(0,3),$ there exists a neighborhood $D(0)$ of $0$ in $\R$ such 
that 
$\big(\sum_{j=1}^{\infty}\big|(\lambda_j^{u_0})^\frac{s}{2}\la\phi_j^{u_0},\cdot\ra_{L^2}\big|^2\big)^{\frac{1} 
{2}}
		\asymp\|\cdot\|_{(s)}$ for any $u_0\in D(0)$.
	\end{remark}

	\begin{lemma}\label{resonances_perturbation}
		Let $B$ satisfy Assumptions I and $N\in\N^*$. Let $\epsilon>0$ small enough and $I^N$ be the set defined 
in 
\eqref{I}. There exists a $D_{\epsilon}\subset\R\setminus\{0\}$ such that, for each $u_0\in D_{\epsilon}$,
		\[
\inf_{\overset{(j,k),(n,m)\in 
I^N}{(j,k)\neq(n,m)}}|\lambda_j^{u_0}-\lambda_k^{u_0}-\lambda_n^{u_0}+\lambda_m^{u_0}|> \epsilon.
\]
		Moreover, for every $\delta>0$ small there exists $\epsilon>0$ such that $dist(D_{\epsilon},0)<\delta.$
	\end{lemma}
	\begin{proof}
		Let us consider the neighborhood $D(0)$ provided by Lemma \ref{invertibility_perturbation}. The maps 
$u\mapsto\lambda_j^{u}-\lambda_k^{u}-\lambda_n^{u}+\lambda_m^{u}$ are analytic for each $j,k,n,m\in\N^*$ and $u\in 
D(0)$. The number of elements such that
		\begin{equation}\label{resonances1}\lambda_j-\lambda_k-\lambda_n+\lambda_m= 0, \ \ \ \ \ \ \ \ \ \  
\forall 
j,n,k,m\in\N^*,\ k,m\leq N\end{equation}
		is finite. Indeed $\lambda_k=k^2\pi^2$ and \eqref{resonances1} corresponds to $j^2-k^2=n^2-m^2$. We have 
$|j^2-n^2|=|k^2-m^2|\leq N^2-1,$ which is satisfied for a finite number of elements.
		Thus, for $I^N$ (defined in \eqref{I}), the following set is finite
		\[
R:=\{((j,k),(n,m))\in (I^N)^2\ :\ (j,k)\neq(n,m); \ \lambda_{j}-\lambda_{k}-\lambda_{n}+\lambda_{m}= 
0\}.
\]
		
		\needspace{3\baselineskip}
		
		\noindent
		{\bf 1)} Let $ ((j,k),(n,m))\in R$, the set $V_{(j,k,n,m)}=\{u\in D\big|\ 
\lambda_j^{u}-\lambda_k^{u}-\lambda_n^{u}+\lambda_m^{u}=0\}$ is a discrete subset of $D(0)$ or equal to $D(0)$. 
Thanks to the relation \eqref{decomposition_perturbed_eigenvalue},
		\begin{equation*}
		\begin{split}
\lambda_j^{u}-\lambda_k^{u}-\lambda_n^{u}+\lambda_m^{u}=&\ 
|a_j|^2\lambda_j+u|a_j|^2B_{j,j}+\lambda_j^{u}\|\eta_j\|_ {
L^2}^2\\
&+u\overline{a_j}\la 
P_{\phi_j}^{\bot}B\phi_j,\eta_j\ra_{L^2}-|a_k|^2\lambda_k-u|a_k|^2B_{k,k}-\lambda_k^{u}\|\eta_k\|_{L^2}
^2\\
&-u\overline{a_k}\la 
P_{\phi_k}^{\bot}B\phi_k,\eta_k\ra_{L^2}-|a_n|^2\lambda_n-u|a_n|^2B_{n,n}-\lambda_n^{u}\|\eta_n\|_{L^2}^2\\
		&-u\overline{a_n}\la 
P_{\phi_n}^{\bot}B\phi_n,\eta_n\ra_{L^2}+|a_m|^2\lambda_m+u|a_m|^2B_{m,m}\\
&+\lambda_m^{u}\|\eta_m\|_{L^2}
^2+u\overline{a_m}\la P_{\phi_m}^{\bot}B\phi_m,\eta_m\ra_{L^2},\\ 
		\end{split}
		\end{equation*}
		which implies
		\begin{equation*}
		\begin{split}
		\lambda_j^{u}-\lambda_k^{u}-\lambda_n^{u}+\lambda_m^{u}
=&\ |a_j|^2\lambda_j-|a_k|^2\lambda_k-|a_n|^2\lambda_n+|a_m|^2\lambda_m\\
		&+\big(|a_j|^2B_{j,j}-|a_k|^2B_{k,k}\\
		&-|a_n|^2B_{n,n}+|a_m|^2B_{m,m}\big)u+o(u).\\ 
		\end{split}
		\end{equation*}
		For $|u|$ small enough, thanks to $lim_{|u|\rightarrow 0}|a_j|^2=1$ and to the third point of Assumptions 
I, each map 
\[
u\mapsto\lambda_j^{u}-\lambda_k^{u}-\lambda_n^{u}+\lambda_m^{u}
\]
can not be constantly equal to $0$. 
Then, $V_{(j,k,n,m)}$ is discrete and $V=\{u\in D\big|\ \exists (j,k,n,m)\in R: 
\lambda_j^{u}-\lambda_k^{u}-\lambda_n^{u}+\lambda_m^{u}=0\}$ is a discrete subset of $D(0)$. As $R$ is a finite 
set $$\widetilde U_{\epsilon}:=\{u\in D:\forall(j,k,n,m)\in R\big|\ 
|\lambda_j^{u}-\lambda_k^{u}-\lambda_n^{u}+\lambda_m^{u}|\geq\epsilon\}$$ has positive measure for $\epsilon>0$ 
small enough. Moreover, for any $\delta>0$ small, there exists $\epsilon_0>0$ such that $dist(0,\widetilde 
U_{\epsilon_0})<\delta.$
		
		\smallskip
		\needspace{3\baselineskip}
		
		\noindent
		{\bf 2)} Let $((j,k),(n,m))\in (I^N)^2\setminus R$ be different numbers. We know 
that $$|\lambda_{j}^0-\lambda_{k}^0-\lambda_{n}^0+\lambda_{m}^0|=\pi^2|j^2-k^2-n^2+m^2|>\pi^2.$$ First, thanks to 
\eqref{decomposition_perturbed_eigenvalue}, we have $\lambda^{u}_j\leq|a_j|^2\lambda_j+|u|C_1$ and 
$\lambda^{u}_j\geq|a_j|^2\lambda_j-|u|C_2$
		for suitable constants $C_1,C_2>0$ non depending on the index $j$. Thus
\[
		|\lambda_{j}^{u}-\lambda_{k}^{u}-\lambda_{n}^{u}+\lambda_{m}^{u}|\geq 
||a_j|^2\lambda_{j}-|a_k|^2\lambda_{k}-|a_n|^2\lambda_{n}+|a_m|^2\lambda_{m}|-|u|(2C_1+2C_2).
\]
		Now, $\lim_{k\rightarrow\infty}|a_k|^2=1$. For any $u$ in $D(0)$ and $\epsilon$ small enough, there exists 
$M_{\epsilon}\in\N^*$ such that, for every $((j,k),(n,m))\in R^C:=(I^N)^2\setminus R$ and $j,k,n,m\geq 
M_{\epsilon}$, $$||a_j|^2\lambda_{j}-|a_k|^2\lambda_{k}-|a_n|^2\lambda_{n}+|a_m|^2\lambda_{m}|\geq 
\pi^2-\epsilon.$$ However 
$\lim_{|u|\rightarrow 0}|a_k|^2=1$ uniformly in $k$ thanks to \eqref{upper_bound_resto} and then there exists a 
neighborhood $ W_{\epsilon}\subseteq D(0)$ such that, for each $u\in W_{\epsilon}$, it follows $$ 
||a_j|^2\lambda_{j}-|a_k|^2\lambda_{k}-|a_n|^2\lambda_{n}+|a_m|^2\lambda_{m}|\geq \pi^2-\epsilon$$ for every 
$((j,k),(n,m))\in R^C$ and $1\leq j,k,n,m < M_{\epsilon}$. Thus, for each $u\in W_{\epsilon}$ and 
$((j,k),(n,m))\in R^C$ such that $(j, k)\neq( n,m)$, we have 
$|\lambda_{j}^{u}-\lambda_{k}^{u}-\lambda_{n}^{u}+\lambda_{m}^{u}|\geq \pi^2-\epsilon.$
		
		\smallskip
		
		\needspace{3\baselineskip}
		
		\noindent
		{\bf 3)} The proof is achieved since, for $\epsilon_1>0$ small enough, $\widetilde U_{\epsilon_1}\cap 
W_{\epsilon}$ is a non-zero measure subset of $D(0)$. For any $u\in \widetilde U_{\epsilon_1}\cap W_{\epsilon}$ 
and for any $((j,k),(n,m))\in (I^N)^2$ such that $(j,k)\neq(n,m)$, we have 
$|\lambda_{j}^{u}-\lambda_{k}^{u}-\lambda_{n}^{u}+\lambda_{m}^{u}|\geq\min\{\pi^2-\epsilon,\epsilon_1\}.$ 
\qedhere\end{proof}
	
	\begin{remark}\label{remark_general_perturbation}
		Let $B$ satisfy Assumptions I. By using the techniques of the proofs of Lemma 
\ref{perturbation_mixing_eigenspaces} and Lemma 
\ref{resonances_perturbation}, one can ensure the existence of a neighborhood $D_1$ of $u_0$ in $\R$ and $D_2$, a 
countable subset of $\R$ such that, for any $u_0\in D(0):=(D_1\setminus D_2)\setminus\{0\}$, we have:
		\begin{enumerate}
			\item For every $N\in\N^*$, $(j,k),(n,m)\in I^N$ (see \eqref{I}) such that $(j,k)> (n,m)$, 
there holds $\lambda_j^{u_0}-\lambda_k^{u_0}-\lambda_n^{u_0}+\lambda_m^{u_0}\neq 0.$
			\item $B_{j,k}^{u_0}=\la\phi_j^{u_0}(T),B\phi_k^{u_0}(T)\ra_{L^2}\neq 0$ for every $j,k\in\N^*.$
			\item Let $T>0$ and $\epsilon_0>0$. For $|u_0|$ small enough, the neighborhood 
$O^{u_0}_{\epsilon_0,T}$ (defined in \eqref{neigh_perturbated}) contains $O_{\epsilon,T}$ (defined in 
\eqref{neigh_base}) for $\epsilon>0$ sufficiently small.
			\end{enumerate}
	\end{remark}

\end{document}